\documentclass[twoside]{article}
\usepackage[latin1]{inputenc}
\usepackage{t1enc}
\usepackage{a4}
\usepackage{tabularx}
\usepackage{epsf}

\def\la{\mathrel{\mathchoice {\vcenter{\offinterlineskip\halign{\hfil
$\displaystyle##$\hfil\cr<\cr\sim\cr}}}
{\vcenter{\offinterlineskip\halign{\hfil$\textstyle##$\hfil\cr
<\cr\sim\cr}}}
{\vcenter{\offinterlineskip\halign{\hfil$\scriptstyle##$\hfil\cr
<\cr\sim\cr}}}
{\vcenter{\offinterlineskip\halign{\hfil$\scriptscriptstyle##$\hfil\cr
<\cr\sim\cr}}}}}
\def\ga{\mathrel{\mathchoice {\vcenter{\offinterlineskip\halign{\hfil
$\displaystyle##$\hfil\cr>\cr\sim\cr}}}
{\vcenter{\offinterlineskip\halign{\hfil$\textstyle##$\hfil\cr
>\cr\sim\cr}}}
{\vcenter{\offinterlineskip\halign{\hfil$\scriptstyle##$\hfil\cr
>\cr\sim\cr}}}
{\vcenter{\offinterlineskip\halign{\hfil$\scriptscriptstyle##$\hfil\cr
>\cr\sim\cr}}}}}

\def\bref{\vspace{4pt}\noindent\hangindent=10mm}
\begin{document}

\setcounter{figure}{0}
\setcounter{section}{0}
\setcounter{equation}{0}

\begin{center}
{\Large\bf
Complexity in small-scale dwarf spheroidal galaxies}\\
{\bf Ludwig Biermann Award Lecture 2008}\\[0.7cm]

Andreas Koch \\[0.17cm]
Department of Physics \& Astronomy \\
University of Leicester\\
University Road, Leicester, LE1 7RH, UK\\
ak326@astro.le.ac.uk 
\end{center}

\vspace{0.5cm}

\begin{abstract}
\noindent{\it
Our knowledge about the dynamics, the chemical abundances and the evolutionary 
histories of the more luminous dwarf spheroidal (dSph) galaxies is constantly growing. 
However, very little is known about the enrichment of the ultra-faint systems 
recently discovered in large numbers in large sky surveys.   
Current low-resolution spectroscopy and photometric data indicate that these 
galaxies are highly dark matter dominated and predominantly metal poor.
On the other hand, recent 
high-resolution abundance analyses indicate that some dwarf galaxies 
experienced highly inhomogenous chemical enrichment, where star formation 
proceeds locally on small scales.  
In this article, I will review 
the kinematic and chemical abundance information 
of the Milky Way satellite dSphs that is  presently available from low- and high resolution spectroscopy. 
Moreover, some of the most 
peculiar element and inhomogeneous enrichment patterns will be discussed and related
to the question of to what extent the faintest dSph candidates   could 
have contributed to the Galactic halo, compared to more luminous systems.
}
\end{abstract}

\section{Introduction}
Dwarf spheroidals (dSphs) are the most common type of galaxies in the Local Group and generally found in the 
proximity (typically closer than 300 kpc) of larger galaxies like the Milky Way (MW) or the Andromeda galaxy, M31. 
Their very low luminosities ($M_V\ga-14$ mag) and low surface brightnesses ($\mu_V\ga22$ mag\,arcsec$^{-2}$) also render them 
the faintest galaxies known to exist in the universe. They are further characterized by total masses of a few 10$^7$ $M_{\odot}$ 
and a puzzling deficiency of gas, with upper limits on their H\,{\sc I} masses of typically $\la 10^5$$M_{\odot}$.  
This  is far below the values expected from red giant mass loss  even on the time scale of several Gyr 
(see Grebel et al. 2003, and references therein, for a recent review of the properties and possible origins of the more luminous dSphs). 
Moreover, the dSphs are fairly  metal poor systems, with mean metallicities [Fe/H] reaching from ca. $-$1 to $-$2 dex, 
while individual stars are found as metal poor as almost $\sim$$-3$ dex.
 
All nearby dSphs, for which sufficiently deep data are available, the MW and the Magellanic Clouds share a common epoch of ancient star formation (SF)  
within the measurement accuracy (Grebel \& Gallagher 2004). 
Some of the dSphs contain dominant intermediate age populations as well (e.g., Gallart et al. 1999), or even signs of some recent SF. 
In those galaxies that host distinct sub-populations in age and/or metallicity, 
there is evidence of population gradients (Harbeck et al. 2003; Tolstoy et al. 2004; 
Ibata et al. 2006) in the sense that the metal rich and younger populations are more 
centrally concentrated than the old and metal poor ones. This is interpreted  
through  deeper potential wells in their centers, which can retain the gas for longer times, thus allowing for 
prolonged chemical enrichment in these regions.  
This extended enrichment also results in a number of intriguing scaling relations, 
such as the well established metallicity-luminosity relation (Dekel \& Woo 2003; Kirby et al. 2008a). 

Finally, their flat velocity dispersion profiles, their lack of any significant rotation and the lack of a depth extent of the dSphs have 
led to the notion that these galaxies are most likely the most dark matter dominated galaxies known to exist (e.g., Gilmore et al. 2007). 
Their mass to light (M/L) ratios derived under simplified assumptions thus reach up to several hundred in Solar units,  
although the role of Galactic tides in the interpretation of the dSphs'  nature and evolution is still under debate (Kroupa 1997; Read et al. 2006; 
Mu\~noz et al. 2008; Pe\~narrubia et al. 2008a). 
\begin{figure}
\begin{center}
\epsfxsize=1\hsize\epsfbox{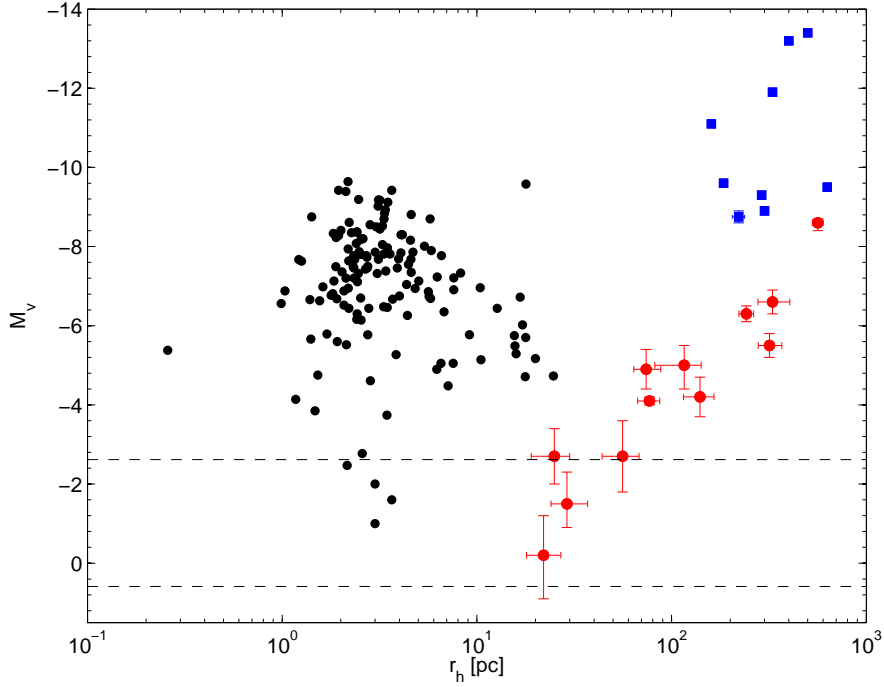}
\end{center}
\caption{Magnitude vs. half-light radius plot for Galactic globular clusters (black points), the 
traditional dSphs (blue squares) and the ultrafaint galaxies (red circles), using data from 
Harris (2003); Gilmore et al. (2007); Koposov et al. (2008); Martin et al. (2008); Koch et al. (2009).  
The dashed lines indicates the luminosity of the tip of the red giant branch for an old (12 Gyr), metal poor ($-$2.3 dex) 
population, using a Dartmouth isochrone (Dotter et al. 2008) 
and the mean empirical red clump magnitude, after Girardi \& Salaris (2001).
}
\end{figure}
In an oft-shown plot of magnitude  versus the galaxies'  radius, however parameterized (Fig.~1; see also Gilmore et al. 2007; 
van den Bergh 2008; Martin et al. 2008), 
the dSphs and the Galactic globular clusters (GCs) clearly separate: while the GCs cut off at sizes of $\la$30 pc, the dSphs exhibit  typical 
radii in excess of $\sim$ 70 kpc. 
An attractive explanation is that the dSphs contain large amounts of dark matter, which forms cored mass distributions 
of a characteristic scale length of the latter size, while the GCs do not (eg., Gilmore et al. 2007).  

The recently discovered {\em ultrafaint} dSphs extend the aforementioned extremes even further, with absolute magnitudes well 
above $M_V > -6$,  and stellar masses in the regime of a mere several thousand to a few ten thousand solar masses.  
Typical core- or half-light radii of these galaxies are of the order of 70 to a few hundred kpc as well (Fig.~1),   
with the exception of a few objects that progressively scatter into the gap and the nature of which is 
yet unclear (see Sect.~1.1). 
Also these galaxies contain old populations of at least 12 Gyr and they are more metal poor on average than 
the previously known, more luminous counterparts; their mean metallicities reach as low as $\sim$$-2.5$ dex (Simon \& Geha 2007).
While no star more metal poor than [Fe/H]$< -3$ dex has been found in any of the classical dSphs (e.g., Koch et al. 2006;2007a,b), 
several such metal poor stars,  
down to $-3.3$ dex  have been detected in the ultrafaint galaxies (Kirby et al. 2008a). 

Deep photometric studies have revealed complex individual properties 
of both the more luminous and the ultrafaint dSphs. Spectroscopy, though sparse, has become 
available for even the fainter stars in the luminous dSphs and in a few of the faintest satellites. 
Yet, the detailed properties of these latter galaxies remain poorly investigated until now. 
How and on what time scales does SF proceed in the faintest, lowest-mass
galaxies? How did galactic winds, outflows or infall of gas, or mixing influence their evolution 
and what role did external effects such as Galactic tides play?  

Cosmological simulations like $\Lambda$CDM predict a wealth of small-scale substructure that gradually merges 
to hierarchically form large-scale structures like present day galaxies such as the MW or M31. 
As a consequence, our Galaxy is expected to be surrounded by 
a large number (hundreds) of dark matter halos. It has been conjectured, whether 
the present-day dSphs could be related to these predicted building 
blocks, but over the years a number of strong arguments against such a simplistic 
view have arisen (Moore et al. 1999). One notion was that  the number of observed satellites 
is too low by a factor 
of several hundred too small compared to the theories. 
This missing satellite discrepancy is nowadays much alleviated, if one accounts for 
the wealth  of newly discovered satellites (Section 1.1)  and efficient mechanisms to suppress SF 
in the dark matter halos at early times, before and during re-ionization (Robertson et a. 2005; Font et al. 2006; Moore et al. 2006; 
Simon \& Geha 2007; Strigari et al. 2007; Tollerud et al. 2008). 
At what redshifts were the ``true'' stellar 
building blocks then accreted 
and how do they relate to the observed population of surviving dSphs; thus: how and when did the 
(ultrafaint) dSphs form and evolve and how do they fit 
into the cosmological $\Lambda$CDM models? In particular, {\em what fraction of dSph-like systems contributed to the build-up of the stellar halo
of the MW?}

In this paper, I will review the most recent results that can be gleaned from {\em spectroscopy} 
both in the traditional, more luminous dSphs and the \mbox{(ultra-)} faint companions to the MW. 
This will reveal a high degree of complexity of these intriguing systems, not only in terms of their 
individual properties and evolution, but also in the context of their r\^ole for our understanding of  
cosmological structure formation. 
I will start by attempting a present census of the traditional and ultrafaint dSphs. 
In Section~2, the kinematic properties of the dSphs will be summarized and a brief account of their 
dark matter properties will be given. Section~3 first focusses on the general metallicity distributions 
of the dwarfs that are predominantly obtained from low-resolution spectroscopy, before discussing  in detail their 
chemical abundances from high-resolution data. 
Section~4 finally summarizes the overall findings with a perspective for future observations. 

\subsection{A census}
The dSphs have always been characterized as very low-luminosity systems, 
with absolute magnitudes fainter than $-$14 mag. For instance, by the time of its discovery 
(Cannon et al. 1977),  the Carina dSph was among the faintest galaxies ever known 
in the Universe ($M_V=-9.3$). 
Thirty years later, the advent of large-scale sky surveys like the Sloan Digital Sky Survey 
(SDSS; Stoughton et al. 2002) or ambitious wide field surveys 
using, e.g., CTIO's MegaCam (Martin et al. 2006), the INT/WFC instrument (Irwin et al. 2007)
or the KPNO Mosaic imager (Majewski et al. 2007),  
has led to the discovery of a vast number of even fainter dSph satellites to the MW 
(Willman et al. 2005a,b; Belokurov et al. 2006a, 2007a; Zucker et al. 2006a,b; Walsh et al. 2007;  
Grillmair 2008) and M31 (Zucker et  al. 2004, 2007, 2008; Martin et al. 2006; Majewski et al. 
2007; McConnachie et al. 2008), thereby tracing the galaxy luminosity function further down to the {\em 
ultra}faint end (e.g., Koposov et al. 2008). 
Table~1 lists the main properties of the currently known dSph candidates of the MW system that are relevant for this review. 
\begin{table}
\begin{center}
\begin{tabular}{rcccrrccc}
\hline
& D$_{\odot}$ & $r_h$ & & $<$v$_{\rm rad}$$>$  &  $\sigma$ & 
 &  & (M/L)$_V$  \\
\raisebox{1.5ex}[-1.5ex]{dSph} &[kpc] & [pc] &  \raisebox{1.5ex}[-1.5ex]{M$_V$} & 
[km\,s$^{-1}$] &  [km\,s$^{-1}$] & \raisebox{1.5ex}[-1.5ex]{$<$[Fe/H]$>$}
&  \raisebox{1.5ex}[-1.5ex]{$\sigma$[Fe/H]} & [(M/L)$_{\odot}$] \\
\hline
Sagittarius          & 28$\pm$3           & $\ga$500    & $-$13.4 & 149.4$\pm$0.6 & 9.6$\pm$0.4 & $-$0.5 & 0.8 & $\sim$22 \\
Fornax	     	  & 138$\pm$8         & 400  		& $-$13.2	     	 &     55.2$\pm$0.1 	    & 11.7$\pm$0.9 & $-$1.3 & 0.5   &	14.8$\pm$8.3	     \\
Leo I	     	  & 250$\pm$30        & 330  		& $-$11.9	     	 &    284.2$\pm$1.0 	    &  9.9$\pm$1.5 & $-$1.3 & 0.2   &	23.5$\pm$4.5	     \\
Sculptor     	  &  79$\pm$4         & 160  		& $-$11.1	     	 &    111.4$\pm$0.1 	    &  9.2$\pm$1.1 & $-$1.5 & 0.5   &  158$\pm$33	     \\
Leo II	      	  & 205$\pm$12        & 185  		& $-$9.6	     	 &     79.1$\pm$0.6 	    &  6.6$\pm$1.5 & $-$1.7 & 0.2   &	33$\pm$5.75	     \\
Sextans	     	  &  86$\pm$4         & 630  		& $-$9.5	     	 &    224.3$\pm$0.1 	    &  7.9$\pm$1.3 & $-$1.9 & 0.4   &	70$\pm$10	     \\
Carina	     	  & 101$\pm$5         & 290  		& $-$9.3	     	 &    222.9$\pm$0.1 	    &  6.6$\pm$1.2 & $-$1.7 & 0.3   &  116$\pm$24	     \\
Ursa Minor   	  &  66$\pm$3         & 300  		& $-$8.9	     	 & $-$245.2$^{+1.0}_{-0.6}$ & 12.0  	   & $-$1.9 & 0.7   &  275$\pm$35 	     \\
Draco 	     	  &  76$\pm$5         & 221$\pm$16	& $-$8.8                 & $-$290.7$^{+1.2}_{-0.6}$ & 13.0  	   & $-$2.0 & 0.7   &  290$\pm$60            \\   
\hline
Canes Venatici I  & 218$\pm$10        & 564$\pm$36	& $-$8.6$\pm0.2$	 &     30.9$\pm$2.0 &  7.6$\pm$0.4 & $-$2.1 & 0.5   &   221$\pm$108	       \\
Hercules	  & 132$\pm$12        & 330$^{+75}_{-52}$ & $-$6.6$\pm0.3$	 &     45.0$\pm$1.1 &  5.1$\pm$0.9 & $-$2.6 & 0.5   &   332$\pm$221	       \\
Bo\"otes I	  &  66$\pm$3         & 242$^{+22}_{-20}$ & $-$6.3$\pm$0.3	 &     99.0$\pm$2.1 &  6.5$^{+2.0}_{-1.4}$ & $-$2.1,$-$2.5  & 0.3 &  680$\pm$275 \\
Ursa Major I 	  &  97$\pm$4         & 318$^{+50}_{-39}$ & $-$5.5$\pm0.3$	 &  $-$55.3$\pm$1.4 &  7.6$\pm$1.0 & $-$2.3 & 0.5   &   1024$\pm$636	       \\
Leo IV	     	  & 160$^{+15}_{-14}$ & 116$^{+26}_{-34}$ & $-$5.0$\pm0.6$	 &    132.3$\pm$1.4 &  3.3$\pm$1.7 & $-$2.6 & 0.8   &	151$\pm$177	       \\
Canes Venatici II & 160$^{+4}_{-5}$   & 74$^{+14}_{-10}$& $-$4.9$\pm0.5$	 & $-$128.9$\pm$1.2 &  4.6$\pm$1.0 & $-$2.2 & 0.6   &	336$\pm$240	       \\
Ursa Major II	  &  30$\pm$5         & 140$\pm$25	& $-$4.2$\pm0.5$	 & $-$116.5$\pm$1.9 &  6.7$\pm$1.4 & $-$2.4 & 0.6   &   1722$\pm$1226	       \\
Coma Berenices    &  44$\pm$4 	      & 77$\pm$10	& $-$4.1$\pm0.5$         &     98.1$\pm$0.9 &  4.6$\pm$0.8 & $-$2.6 & 0.5   &	448$\pm$297	       \\
Bo\"otes II       &  46$\pm$4         & 56$\pm$12	& $-$2.7$\pm$0.9	 & $-$117.0$\pm$5.2 & 10.5$\pm$7.4 & $-$1.8 & 0.1   &	 \dots  	       \\
Willman I         &  38$\pm$7         & 25$^{+5}_{-6}$	& $-$2.7$\pm$0.7	 &  $-$12.3$\pm$2.5 &  4.3$^{+2.3}_{-1.3}$ & $-$1.5 & 0.4   &	$\sim$500	       \\
Segue 1	    	  &  23$\pm$1         & 29$^{+8}_{-5}$	& $-$1.5$^{+0.6}_{-0.8}$ &    206.0$\pm$1.2 &  4.3$\pm$1.3 & $-$3.3 & \dots & 1320$\pm$2680	       \\
SDSS J1058+2843   &  24$^{+3}_{-2}$   & 22$^{+5}_{-4}$	& $-$0.2$^{+1.1}_{-1.0}$ & \dots            & \dots        & \dots  & \dots &		\dots	       \\
Leo T	       	  &  407$\pm$38       & 178$\pm$39      & \dots			 &     38.1$\pm$2.0 &  7.5$\pm$1.6 & $-$2.0 & 0.5   &  138$\pm$71 	       \\
Bo\"otes III   	  &  46$\pm$1         & $\sim$1000      & \dots  		 & \dots            & \dots        & \dots  & \dots & \dots   		       \\
\hline
\hline
\end{tabular}
\caption{Properties of the classical and the ultrafaint dSphs relevant for this review: 
(1) Name; (2) heliocentric distance; (3) half-light radius; (4) absolute V-band magnitude; (5,6) radial velocity and overall velocity 
dispersion; (7,8) mean metallicity and 1$\sigma$-spread; (9) V-band mass-to-light ratio. 
Sources for the data are Mateo (1998); Grebel et al. (2003); Wilkinson et al. (2004); Koch et al. (2006, 2007a, 2007b); 
Gilmore et al. (2007); Simon \& Geha (2007); Bellazzini et al. (2008); Martin et al. (2008); Walker et al. (2009); and references therein. 
Values for Bo\"otes II, III and Segue 1 were adopted from Koch et al. (2009); Grillmair (2008) and Geha et al. (2008); while 
metallicity measurements for the remaining ultrafaint galaxies are from  Martin et al. (2007); Kirby et al. (2008a). 
Spectroscopic metallicities on the scale of Carretta \& Gratton (1997)  were adopted where available. For completeness 
I list both deviating [Fe/H] measurements from the spectroscopic studies of Mu\~noz et al. (2006) and Martin et al. (2007) 
for Bo\"otes~I. Listed velocities and dispersions do not distinguish between claims of kinematical substructures, if present 
(e.g., CVn~I; Ibata et al. 2006). Although listed, the disrupting Sagittarius dSph deviates from 
all stated relations and is excluded from the discussions in this review.}
\end{center}
\end{table}
As Fig.~1 implies, many of these new satellites are comparable in magnitude to the (Galactic) GCs, while their spatial extent can extend to up to two orders of magnitude higher. 
The resulting very low surface brightnesses render it obvious that they have been so elusive from 
past shallower sky surveys. 

And yet the terminology regarding this new generation of satellites is often misleading: In the literature,  
``ultrafaint'' is generally applied to those dSphs recently discovered in the 
SDSS or other sky surveys.  One should keep in mind, though, that some of these 
objects are still relatively bright; for instance CVn~I (Zucker et al. 2006b) is as luminous as the ``traditional'' dSph Draco,  and with a half light radius of $\sim$560 pc also the most extended 
MW dSph, while the Hercules dwarf (Belokurov et al. 2007a), at $M_V=-6.6$, 
has about one tenth of Draco's luminosity. 
Thus, for a clear distinction, the label ``ultrafaint dSph'' should strictly be applied to systems fainter than a magnitude 
cut off at $M_V \ga -6$ mag. 
Throughout this review, the more luminous dSphs, meaning those known in the pre-SDSS era, will also be referred to as the ``traditional'' or ``classical'' dSphs. 

A number of the faintest galaxies have total absolute magnitudes that are comparable or even fainter than  
the absolute magnitude of the tip of (theoretical) red giant branches (RGBs; dashed line in Fig.~1) 
and their color magnitude diagrams (CMDs) show only a handful of evolved stars. At such low magnitudes, these galaxy-contenders 
are reminiscent of the faintest, peculiar halo clusters (AM~4, Palomar~1, Kop~1, 2; Inman \& Carney 1987; 
Rosenberg et al. 1998; Koposov et al. 2007) that stand out 
through their absence of any significant RGB, although the dSph radii are larger by a factor of up to ten. 
Considering this  sparsity in color-magnitude space and the fact that a few of these objects   
occupy the gap or transition region in the magnitude vs. radius plot (Fig.~1), the actual nature of the faintest 
stellar overdensities  (Segue~1, Willman~1, Bo\"otes~II) remains unclear (see also Liu et al. 2008). In particular,  
their relatively small radii have prompted suggestions that these systems may be inflated star clusters, dense parts of tidal streams 
or heavily tidally stripped dwarf galaxies rather than classical old and metal poor dSphs. 
Other interpretations argue that these ultrafaint objects could be the stripped remnants 
from larger satellites such as the disrupted Sagittarius system (Ibata et al. 1994; Koch et al. 2009). 
 
At present, there are nine luminous dSph satellites known to belong to the MW and of the order of 12--15 
faint to ultrafaint satellites, modulo the aforementioned uncertainties in the classification of 
some of the faintest candidates. The count of M31 satellites, on the other hand, has reached as 
far as And~XX (McConnachie et al. 2008), although it should be noted that And~IV  
turned out to be a background galaxy (e.g., Ferguson et al. 2000), and And~VIII  is likely a tidal disrupted system 
and associated with M31 halo streams (Morrison et al. 2003). 
In either case, the discovery rate is proceeding at a fast pace and a wealth of new such objects is expected in the near future. 

\section{Kinematics}
Radial velocities of the dSphs are usually measured from the Doppler shifts of prominent absorption features 
in red giant spectra. Depending on the spectrographs' wavelength coverage and  the achieved signal-to-noise 
ratios, the most common spectral reference features are the magnesium triplet lines at $\sim$5150\AA\ (e.g., Walker et al. 2007) and 
the near-infrared calcium triplet (CaT) lines at $\sim$8500\AA\ (e.g., Kleyna et al. 2002; Koch et al. 2007a,c; see also Fig.~3).  
Nowadays individual radial velocities in dSphs are published for several thousand stars in all of the more luminous 
galaxies, e.g.,  Fornax (Walker et al. 2009) or Carina (Koch et al. 2006; Mu\~noz et al. 2006a; Walker et al. 2009),  
a few hundred in the classical Sextans, Draco, Ursa Minor and Sculptor (Kleyna et al. 2004; Wilkinson et al. 2004; 
Battaglia et al. 2008a; Walker et al. 2007,2009) and  
of the order of 50--200 in the fainter and/or remote dSphs like Bo\"otes~I, CVn~I, Leo~I and II (Mu\~noz et al. 2006b; 
Ibata et al. 2006; Koch et al. 2006a,c). 
Reassuringly, present-day accurate radial velocity dispersion {\em profiles} 
from the numerous data sets have practically confirmed mass estimates from earliest measured central dispersion values 
(e.g., Aaronson 1983). 

Typically, the stellar velocity dispersions in the traditional dSphs are of the order of 10 km\,s$^{-1}$; 
their radial velocity dispersion profiles remain essentially constant out to the last observed data points. 
In fact, most of the dSph profiles have been mapped out to the large radii, typically a few tens of arc minutes (see also Fig.~1),  
at which their surface brightnesses level off into the background\footnote{Under a common misconception, this is often 
paraphrased as the ``tidal'' radius, although a simple term as ``stellar limiting radius'' seems 
more appropriate in the context of tides and the dSphs (see, e.g., discussions in Koch et al. 2007a; Gilmore et al. 2007).}. 
Given the low stellar densities in the outer parts, however, the outer radial bin in the profiles usually contains only 
few stars, leading to larger uncertainties for these  outermost data points.
Deviations from a flat dispersion profile have been reported for individual cases, such as Ursa Minor (Wilkinson et al. 2004) 
with a significantly colder population in the outer parts,  or Leo~I (Koch et al. 2007a), which shows an indication of 
a rising profile in the innermost parts (cf. Sohn et al. 2007; Walker et al. 2007). Yet all the observed profiles 
are statistically consistent with simple, single-population mass models (Gilmore et al. 2007), 
without the need to invoke superpositions of multiple stellar populations with different dispersions or scale lengths 
(e.g., McConnachie et al. 2006).

In the ultrafaint satellites the RGB becomes progressively fainter and sparser and 
much of the observing time is spent of vetting Galactic  foreground contamination, unless methods are 
used for target selection that permit dwarf/giant separation, such as Str\"omgren photometry (e.g., 
Faria et al. 2007), Washington photometry (e.g., Majewski et al. 2000) or photometric combinations (e.g., Koch et al. 2008c).  
This aggravates the measurability of accurate velocity distributions and mostly inhibits 
determinations of {\em spatially resolved}, i.e., 
radial, velocity dispersion profiles. Typically, a handful to a few dozens of member stars are measured within each  
of the ultrafaint galaxies (e.g., Simon \& Geha 2007; Geha et al. 2008; Koch et al. 2009). 
As a result,  the ultrafaint dSphs  exhibit lower velocity dispersions on average, typically in the range of 3--7 km\,s$^{-1}$ 
(Table~1).  

Coupled with the declining surface brightness profiles in the dSphs, their flat radial dispersion profiles are 
inconsistent with simple mass-follows-light models, but rather indicate that 
the dSphs are embedded in massive {\em dark matter} halos. 
Estimates of their total masses from the central dispersion, $\sigma$, which scales as  
 $M_{\rm tot}\propto r_c\,\sigma^2$ (with core radius $r_c$; King 1996; Illingworth 1976) 
already imply masses for the traditional dwarfs of a few times $10^7$ M$_{\odot}$ and of the order of a few times $10^6$ 
M$_{\odot}$  for the ultrafaint dSphs. Coupled with the very low luminosities of these galaxies, this 
further indicates high mass-to-light (M/L) ratios of up to several hundred in Solar units (see also Mateo 1998; 
Table~1; and references therein).  These extreme values have led to early conjectures that the dSphs are 
likely the most dark matter dominated objects to exist on small scales. 
Alternative views to interpret the observed dynamics of the dSphs, however, leave room  for modified gravity 
theories (MOND; e.g., {\L}okas 2001,2002) 
or to identify them as tidal remnants (e.g., Kroupa 1997), both without the need to invoke dark matter. 
In particular the latter is, however, at odds with the lack of a depth extent of the dSphs, as shown by Klessen et al. (2003). 

Detailed mass {\em profiles} are then derived from the radial velocity dispersion profiles. The most straightforward approach 
applies the Jeans equations to relate the underlying total mass distribution to the observed brightness profiles and kinematics 
of the tracers, viz. the red giants (Binney \& Tremaine 1987), although sophisticated nonparametric models are progressively 
developed (e.g., Wang et al. 2005; Strigari et al. 2007).  
Although widely applied, the mass determinations via the Jeans equations should be taken with caution, given the  
number of simplifications that enter the modeling. Amongst these are the assumption of spherical symmetry, velocity isotropy 
and the assumption of dynamical equilibrium, that is, neglect of Galactic tides. That these are mostly too simplistic is 
illustrated, e.g., by the cases of  Hercules, which shows  an unusually large elongation that may be of tidal origin
 (Coleman et al. 2007; Martin et al. 2008), Ursa Maior II with its irregular shape (Zucker et al. 2006a; Belokurov et al.
 2007b; Fellhauer et al. 2007) and relatively high velocity dispersion (Martin et al. 2007; Simon \& Geha 2007),  
and the prime example of Sagittarius, which is clearly undergoing tidal disruption during its accretion onto the MW (Ibata et al. 
1994). 
Moreover, the anisotropy parameter in the models 
 is degenerate with the shape of the (inner) density profile;  the fits of present-day velocity dispersion profiles 
are unable to statistically  differentiate between constant or radially varying velocity anisotropy (e.g., {\L}okas 2002; 
Koch et al. 2007a,c; Battaglia et al. 2008a), nor can they conclusively distinguish between cored or cusped  
density profiles unless higher moments of the velocity distributions are taken into account (e.g., {\L}okas 2009). 
Although purely kinematic data are not yet sufficient to resolve the controversy of cosmologically motivated 
cusped (Navarro et al. 1997) versus empirical, cored density profiles 
(e.g., Hernquist 1990; {\L}okas 2002; Kleyna et al. 2003; Read \& Gilmore 2005; Sanchez-Salcedo et al. 2006), 
other pieces of evidence favor cored inner mass distributions. In particular, phase-space substructures 
in some dSphs resembling star clusters (Kleyna et al 2003) or even an intact GC system like in Fornax (Goerdt et al. 2006) 
would have quickly dispersed in the presence of  an inner density cusp.
Also,  in cases where tentative evidence of two or more subpopulations may be present, whether distinct in kinematics, 
metallicity and/or age (as, e.g., claimed for Sculptor or CVn~I; Tolstoy et al. 2004; Ibata et al. 2006), 
single-population Jeans modeling  may not yield accurate results (McConnachie et al. 2006; Battaglia et al. 2008a). 

On the other hand, apart from the few clear cases discussed above that are in a state of tidal disruption, 
dSphs are pressure supported systems in which no net rotation has been detected to date (Koch et al. 2007a,c; Gilmore 
et al. 2007; but see also Sohn et al. 2007; Mu\~noz et al. 2008;  for an alternate view), which again argues in favor of them being 
dark matter dominated objects. However, any significant tidal stirring would predominantly act only at large radii that 
are mostly outside the presently targeted areas  (e.g., Pe\~narrubia et al. 2008b). 
Another important notion regarding Galactic tides, if acting, is that they efficiently remove mass from the dSphs over their 
life times (e.g., Read et al. 2006; Mu\~noz et al. 2008). Thus the observed total mass and mass profiles 
are not necessarily representative of the initial mass, as usually nothing is known about their mass loss {\em history}. 
This permeates the interpretation of the dSphs' whole evolutionary histories: 
If a galaxy started with a much larger 
initial (stellar) mass, it will have experienced a much different (viz. prolonged) chemical enrichment. 
Thus such heavily affected systems would seem too metal rich for their present day mass or luminosity so that they tend to deviate 
from well-defined scaling relations like the metallicity-luminosity relationship. 
This is in fact seen in a few of the ultrafaint candidates such as Bo\"otes~II (Koch et al. 2009; see also Sect.~3.1). 

One of the earliest notions on the masses of the dSphs was that of Mateo et al. (1993) and Mateo (1998), 
namely that M/L scales with the galaxies' luminosity.
This was confirmed with the new and better measurements of  the luminous dSphs 
and expanded to the (ultra-) faint regime. Fig.~2 shows an updated version of 
the Mateo-plot using the presently available data. 
\begin{figure}
\begin{center}
\epsfxsize=1\hsize\epsfbox{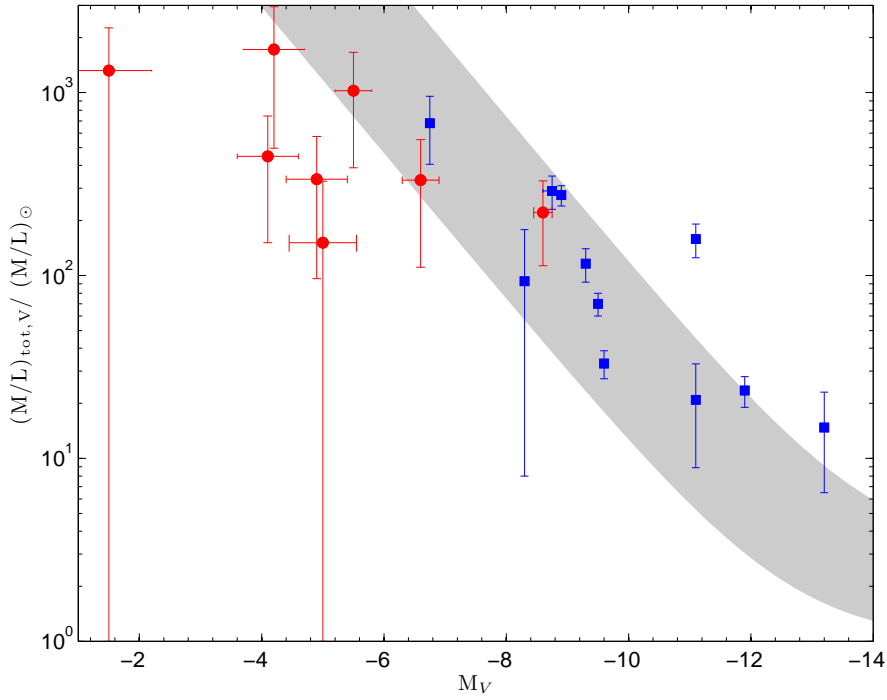}
\end{center}
\caption{Mateo-plot of mass to light ratio vs. absolute magnitude of the traditional, more luminous dSphs 
(blue squares) and the ultrafaint satellites (red symbols), using data from C\^ot\'e et al. (1999); Wilkinson 
et al. (2004); Kleyna et al. (2005); Chapman et al. (2005); Wang et al. (2005); Wilkinson et al. (2006); 
Mu\~noz et al. (2006); Koch et al. (2007a,c); Simon \& Geha (2007); Battaglia et al. (2008a); 
Martin et al. (2008); Geha et al. 
(2008). The gray shaded area indicates the parameter space covered assuming a combination of 
stellar M/L from 1--3 and a constant dark matter halo mass range of 1--10$\times 10^7$ M$_{\odot}$ 
(cf.  Strigari et al. 2008).  
The value obtained for Sculptor (at $M_V=-11.1$), under the assumption of two distinct populations, clearly 
deviates from the trend outlined by the more luminous galaxies.}
\end{figure}
The narrow trend  that is outlined by the more luminous satellites implies a common underlying dark matter halo mass scale,  
in which  the dSphs are embedded. Depending on an assumed stellar M/L (usually chosen as 1--3 (M/L)$_{\odot}$), 
the present data are consistent with a halo mass of a few times $10^7$ M$_{\odot}$ (gray shaded area in Fig.~2).  
The ultrafaint galaxies, however, deviate from the trend in Fig.~2 and Simon \& Geha (2007) argue that those systems 
are rather embedded in dark matter halos of a smaller mass. That is, if there was a physical minimum mass scale 
for dark matter halos, it is smaller than those scales sampled by the present-day luminous- and ultrafaint dSph 
data. 
Walker et al. (2007) find that the mass within a radius of 600 pc lies in this limited range for those luminous dSphs in their study, 
while  Strigari et al. (2008) suggest that such a common mass scale is well manifested 
in the galactic mass integrated within 300 pc and found both for the traditional, more luminous dSphs {\em and} the 
ultrafaint satellites. 
In either case, all studies to date suggest an order of magnitude of $\sim10^7$ M$_{\odot}$; 
the existence of such a mass scale at all can be interpreted as due to either the possibility that 
dark matter halos with baryons below this limit simply do not exist, or that star and galaxy formation is suppressed 
in halos below this mass scale. Accurate mass modeling, in particular at the low-luminosity end, 
has to efficiently establish the characteristic clustering scale for dark matter as to ultimately  
constrain the properties of dark matter particles (Gilmore et al. 2007).   

\section{Chemistry}
The chemical element distribution of stars in dSphs is invaluable for studying their chemical enrichment histories: 
while the overall metallicity\footnote{In spectroscopic studies, ``metallicity'' is generally paraphrased as ``iron abundance'' [Fe/H], 
while strictly the ``true metallicity'' [$M$/H] accounts for all heavy elements and has non-negligible contributions 
from the $\alpha$-elements (Section~3.2.1). In the following I will use the notation ``[Fe/H]'' for both terms synonymously.} 
distributions (MDFs)  are well suited to derive the overall, integrated SF and enrichment history of a system, 
knowledge of the detailed chemical element abundance trends is  required to get estimates of the time scales for enrichment 
and to isolate the predominant modes of SF in these low-mass objects. 
\subsection{Metallicities}
Metallicities
 of stars in the faint dSphs are not easily measurable. First estimates of the systems' overall metal 
content are usually obtained by matching sets of theoretical isochrones or empirical fiducials  of Galactic GCs
with known ages and metallicities to the CMDs. While this procedure yields satisfactory results for 
the more luminous galaxies, there are several aggravating factors in the ultrafaint dSphs. 
Firstly, many of the galaxies do not have well populated RGBs and/or are remote and faint so that the age-sensitive 
main sequence turn-offs are generally not discernible. 
Secondly, in those galaxies that also host stellar populations significantly younger than 10 Gyr besides the omnipresent 
old populations, 
age and metallicity are degenerate on the RGB. As a result, metal poor and old tracks occupy the  
same region in the CMD  as the young and metal rich ones (e.g., Koch et al. 2006).
For such cases, stellar spectroscopy is the only viable tool to break undesired degeneracies and 
to obtain accurate metallicity estimates. 
As Fig.~3 indicates, this is well achievable for the brighter red giants in the more luminous dSphs. 
\begin{figure}
\begin{center}
\epsfxsize=0.49\hsize\epsfbox{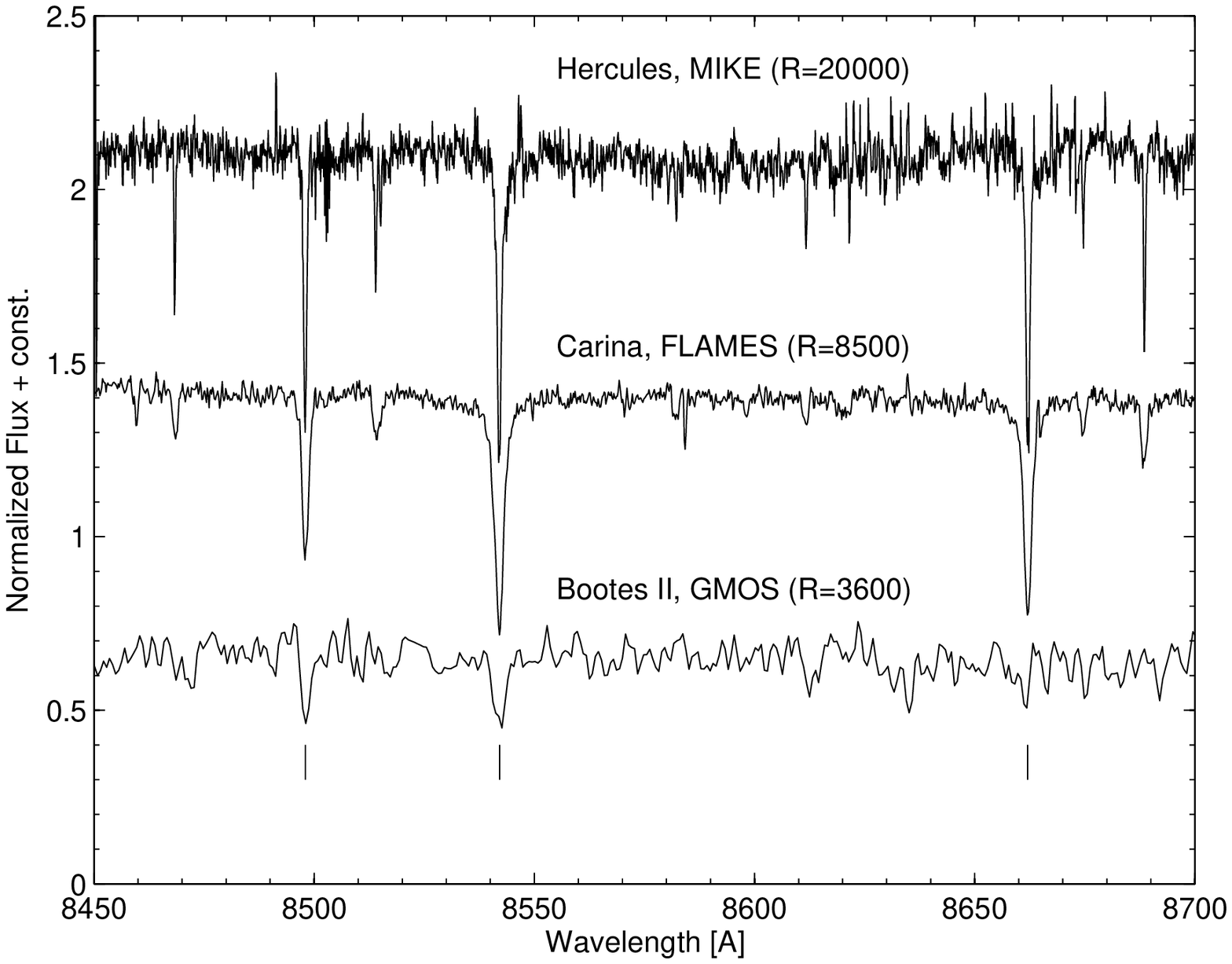}
\epsfxsize=0.49\hsize\epsfbox{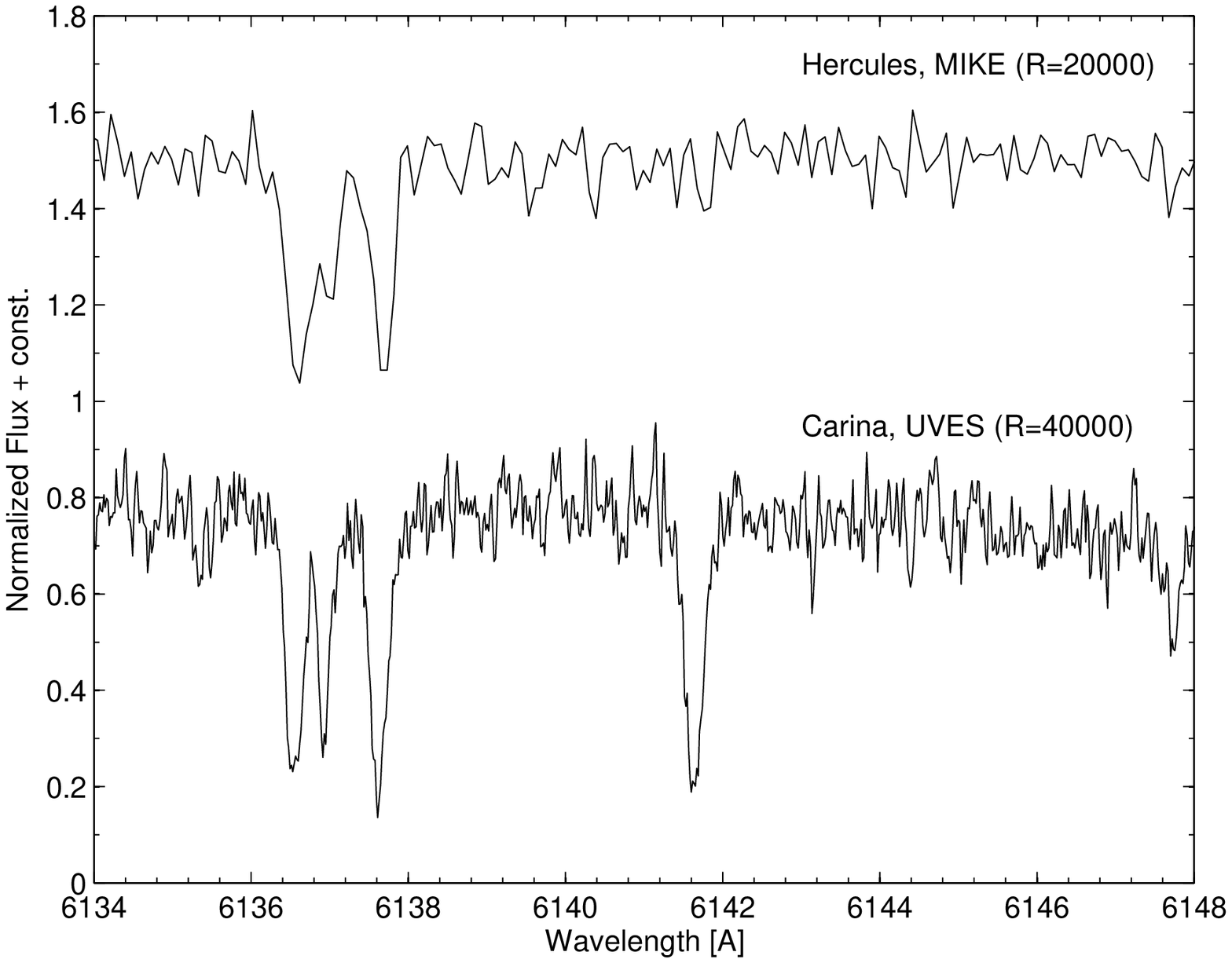}
\end{center}
\caption{Sample spectra covering a representative range of present day's spectrographs'  
resolutions (R) and signal-to-noise ratios. Shown are spectra of  red giants in 
Hercules (V=18.7), Carina (V=17.8), and Bo\"otes~II (V=19.3) around the near-infred CaT 
(indicated as vertical lines; left panel), and high-resolution spectra in Hercules (V=18.7) and Carina (V=17.8) 
around the region of the Ba\,{\sc II}  line at 6141.7\AA\ (right panel). 
The other visible absorption features are  Fe\,{\sc I} lines.}
\end{figure}
On the other hand, this becomes more problematic for the ultrafaint satellites, which either 
do not contain any significant RGB populations that could be easily vetted from foreground stars, 
or which have RGBs that are to faint to be reliably targeted (see also Fig.~1).  
However, recent studies  
have demonstrated that sufficient signal-to-noise ratios for stars 
down to V$\sim$20 mag are achievable employing current state-of-the-art  spectrographs at 8--10 m class telescopes 
(e.g., Koch et al. 2006; Simon \& Geha 2007; Koch et al. 2009) and by applying sophisticated measurement techniques 
(Kirby et al. 2008b; Koch et al. 2008c).  

The traditional dSph galaxies are fairly metal poor system that cover a broad range in mean metallicity, depending 
on their overall luminosity. As Table~1 illustrates, typical values reach from the more rich systems like Fornax (at a mean 
[Fe/H] of $-1.3$ dex) to predominantly metal poor systems like Draco, at $\sim$$-2$ dex. 
The discovery and follow-up photometry of the ultrafaint systems already indicated, however, that these galaxies are even more metal 
poor on average, as for some of them was confirmed from low-resolution spectroscopy. 

Traditional ways to measure the spectroscopic metallicity of dSph red giants are spectral synthesis, using spectral indices such as the 
Mg triplet index (Mu\~noz et al. 2006; Walker et al. 2009) or the wide-spread, empirical calibration of the near-infrared  
CaT onto metallicity [Fe/H]. The latter method has first been established for simple 
stellar populations such as the GCs (Armandroff \& Zinn 1988; Armandroff \& Da Costa 1991; Rutledge et al. 1997a,b). 
Over the past years, however, the calibrations have been successfully applied  to mixed-age populations as the dSphs 
(e.g., Suntzeff et al. 1993; C\^ot\'e et al. 1999; Koch et al. 2006, and references therein). 
In practice, the equivalent widths of the prominent lines of the singly ionized calcium ion at 8498, 8542, 8662\AA\ are correlated with the stellar magnitude above the horizontal 
branch, by which undesired dependences of line strength with effective temperature and surface gravity are removed to 
first order (e.g., Rutledge et al. 1997a; Cole et al. 2000). 
The resulting line index (W$^\prime$) is almost entirely  a function of stellar metallicity and has been accurately 
calibrated onto reference scales using red giants in Galactic GCs of known metallicity (Zinn \& West 1984; Carretta \& Gratton 1997; Kraft \& Ivans 2003). 
The original calibrations were strictly only defined in a limited age and metallicity range, dictated by the GCs 
of the calibration sample. However, recent studies have extended the calibration range towards the metal poor regime 
(Battaglia et al. 2008b), towards metal rich populations (Cole et al. 2004; Carrera et al. 2007) and over 
a broad age range (Cole et al. 2004) through open clusters. 

Despite the straightforward measurement (as strong features in a wavelength region easily accessible using present-day 
instruments),  a number of caveats have emerged in the literature, such as  
the unknown age dependence of the horizontal branch in mixed stellar populations (see discussions in Koch et al. 2006), 
or the vague first-order transition from the 
{\em calcium} line strength to general {\em metal}, or [Fe/H],  abundance:  
by calibrating dSph stars onto a Galactic GC scale one strictly presupposes that the [Ca/Fe] ratio in the dSphs 
is the same as in the Galactic calibrators, while the abundance ratios in the dSphs are either {\em a priori} 
unknown or depleted with respect to the GC stars by up to 0.4 dex  (Bosler et al. 2007; Koch et al. 2008a; Section~3.2.1). 
Reassuringly,  stars that have both low-resolution CaT metallicities and high-resolution iron abundances available agree well 
to within the uncertainties. Systematic deviations (of the order of 0.1 dex) occur only above $-1.2$ and more metal poor 
than $-2.2$ dex (Battaglia et al. 2008b; Koch et al. 2008a).  
In any case the CaT method  has proven a useful tool to rank the galaxies' metallicities and to construct their overall 
MDFs.

Spectroscopic MDFs now exist for almost all of the Galactic satellites:   
Fornax (Pont et al. 2003, Battaglia et al. 2006); 
Leo~I (Bosler et al. 2007; Koch et al. 2007a); 
Sculptor ( Tolstoy et al. 2004; Battaglia et al. 2008b);
Leo~II (Bosler et al. 2007; Koch et al. 2007b); 
Sextans (Helmi et al. 2006); 
Carina (Koch et al. 2006);
and Mu\~noz et al. (2006); Ibata et al. (2006); Martin et al. (2007); Simon \& Geha (2007); Kirby et al. (2008a); Koch et al. (2009) 
for the most recently detected faint to ultrafaint dSphs. 
Although analysed in detail in high-resolution abundance studies (see Sect.~3.2; and references therein) 
MDFs for the classical Ursa Minor and Draco dwarf galaxies are only available from broad- and narrow band photometry 
(e.g., Bellazzini et al. 2002; Faria et al. 2007). 
In Fig.~4 I show exemplary MDFs of four  Galactic satellites, covering a wide range in luminosities. 
\begin{figure}
\begin{center}
\epsfxsize=1\hsize\epsfbox{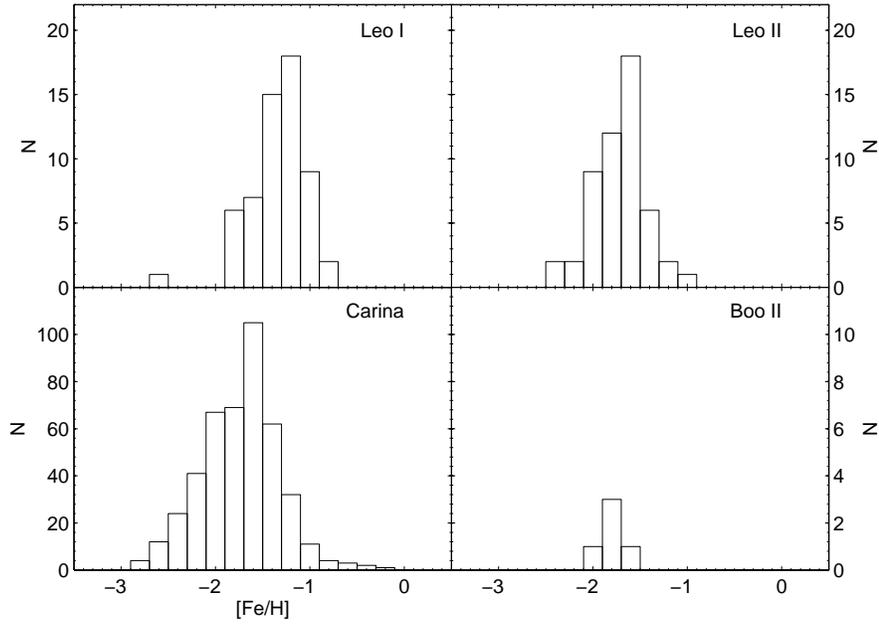}
\end{center}
\caption{Metallicity distributions for four Local Group dSphs, with $M_V$ increasing left to right, 
top to bottom. 
Data are from Koch et al. (2006;2007a,b; 2009). All distributions are based on the CaT
and on the scale of Carretta \& Gratton (1997).}
\end{figure}

While each individual dSph has experienced a unique SF history and no two dSphs are alike 
(Grebel 1997), there are important similarities in the overall shapes of their    
MDFs (e.g., Koch et al. 2007b) and I note the following characteristics:

{\em (1) Metal poor mean:} 
The low values of the mean [Fe/H] indicate a low SF efficiency in the dSphs with only a small  
supernovae (SNe) rate to enrich the interstellar medium (ISM) out of which the subsequent generations 
of stars are born (e.g., Lanfranchi \& Matteucci 2004; see also Sect.~3.2.1). 
A problem with the low SF rates is, though, that these leave a reservoir of gas at the end of the last epoch of SF in the model 
predictions. To explain the observed  gas deficiency in all dSphs  efficient gas removal mechanisms have then to be invoked, 
such as tidal stripping (Ikuta \& Arimoto 2002; Gallagher et al. 2003; Grebel et al. 2003). 

As already indicated by the slight shift in the MDF peaks in Fig.~4, dSphs follow a well defined 
luminosity-metallicity relation (Dekel \& Woo 2003; Grebel et al. 2003): since the more luminous galaxies have deeper 
potential wells,  
they retain their metals for longer time scales, thus 
allowing for stronger or prolonged enrichment (e.g., Dekel \& Silk 1986). 
Given the larger uncertainties on the ultrafaint dSph MDFs and their very broad spreads (see item 2) it is yet  
unclear, whether this relationship extends to the least luminous galaxies (Kirby et al. 2008a) or whether one actually 
observes a leveling off above $M_V\ga-7$ (Simon \& Geha 2007). 
On average, all ultrafaints measured so far have mean metallicities below $-$2 dex, with two exceptions: Firstly, Willman~1 
is possibly a  star cluster (Willman et al. 2005b). Secondly,  Bo\"otes~II may be a heavily stripped object or 
a star cluster from the Sgr system (Fig.~4; Koch et al. 2009). For both cases, the present-day observed mass and luminosities would 
be much smaller than in their initial state, which would be more representative of the observed higher metallicities. 

In the models of Salvadori \& Ferrara (2008), the gas paucity and very low SNe rate in the ultrafaint dSphs essentially 
results in their  evolving as closed boxes, 
which does not hold for the luminous ones (Helmi et al. 2006; Koch et al. 2006, 2007b; see item 3). 
As a consequence these galaxies would retain their metals and get enriched to ``higher'' metallicities than 
expected from the canonical luminosity-metallicity relation. 
Their  still comparably low mean metallicites below $-$2 dex would then be explicable if these ultrafaint systems formed 
very early on, at a time when the available gas was not yet enriched to the higher values. This would require 
a formation well before re-ionization ($z>8.5$; Simon \& Geha 2007; Salvadori \& Ferrara 2008). 

{\em (2) Broad metallicity range:} 
Another important notion is that the dSphs have very broad abundance spreads with formal 1$\sigma$ widths 
(that is, after correction for broadening through measurement uncertainties) of approximately 0.5 dex and 
large spreads of up to 0.8 dex in the ultrafaint ones. The whole range covered in a given dSph is, however, 
very large and usually at least 1 dex wide (Fig.~4). In particular cases like the Carina dSph, stars 
are found covering a full range from nearly $-$3 dex up to near-Solar (Koch et al. 2006). 
Carina is a special case, though, since it has experienced episodic SF (Smecker-Hane et al. 1994; 
Monelli et al. 2003) that led to the occurrence of multiple stellar populations distinct in metallicity and thus an 
efficient broadening of the MDF. Other contenders that were claimed to host populations distinct in spatial 
distribution, kinematics {\em and} metallicity are Sculptor (Tolstoy et al. 2004) and CVn~I (Ibata et al. 2006). 

One possible explanation for the broad metallicity ranges despite the low luminosities is again that 
the dSphs contain large amounts of dark matter (Sect.~2) so that they retained their metals over a long period, leading to a broad 
range in enrichment. This view is very simplistic, though, as nothing is known about the mass loss history of the  
dSphs and thus their initial masses. It is just as conceivable that they started as systems with much higher stellar mass than 
observed today, which was progressively lost -- masses that could lead to the same amount of enrichment.   
Alternatively, the broad ranges could be explained by highly localized, highly inhomogeneous enrichment (e.g., 
Marcolini et al. 2006). 

{\em (3) Metal rich tail:} 
A feature seen in the MDFs and predicted by models is that the distributions cut off more sharply at the metal rich end 
compared to the metal poor tail. This indicates the occurrence of strong outflows in the form of galactic winds, 
typically several times the SF rate (Lanfranchi \& Matteucci 2004, 2007). These strong and continuous 
winds efficiently drive out metals, preventing further enrichment towards the metal rich tail. 
In general, galactic winds and gas outflows play an important role in the chemical evolution of the dSphs and in 
shaping their MDFs (e.g., Mac Low \& Ferrara 1999; Hensler et al. 2004).  

{\em (4) Lack of metal poor stars:} 
Finally, I note the lack of any stars more metal poor than $-$3 dex  in all of the the {\em more luminous dSphs} studied to date. 
That is, these galaxies suffer from a pronounced G-dwarf problem, or, given the evolved nature of the targeted stars, a 
K-giant problem (Shetrone et al. 2001, 2003; Koch et al. 2006; 2007a,b; Helmi et al. 2006).  
Targets for measuring MDFs are generally selected from large samples of often several hundred stars 
to cover the full RGB color range without any observational bias or 
{\em a priori} knowledge of their metallicities -- in that way one ensures to include potential very metal rich or 
extremely metal poor stars, if present. 
The fact that still no very metal poor stars below $-$3 dex are found then indicates that this appears to be a real absence 
in the luminous dSphs. Thus it seems that SF and enrichment in these systems clearly proceeds differently from, e.g., 
the Galactic halo, which contains a few handful of such extremely metal poor stars from $-$5 to $-$3 dex, though incompletely 
sampled  (e.g., McWilliam et al. 1995; Beers \& Christlieb 2005; Cohen et al. 2008). 
As a consequence it can be ruled out that these dSphs evolved according to a closed box scenario, but it is 
rather accepted that they have experienced an early ``prompt'' pre-enrichment (e.g., Tinsley 1975), leading to an initial 
non-zero metallicity. The question of the origin of such pre-enrichment of the initial gas phase is still under debate, 
but a likely scenario is an early enrichment to higher metallicities by pregalactic Population~III stars (e.g., 
Larson 1998; Schneider et al. 2002; Bromm \& Larson 2004). However, this leaves the question of why then 
there are extremely metal-poor stars present in the Galactic halo. 

Interestingly, there is recent evidence that the ultrafaint dSphs may in fact host stars as low as $-3.3$ 
dex with a distribution that resembles that of the Galactic halo (Kirby et al. 2008b).  This is an important 
finding as it reinforces the idea envisioned in the original hierarchical formation scenario (Searle \& Zinn 1978).  
Thus the {\em metal poor halo} of the Galaxy could have been donated by 
dissolving objects like the ultrafaint dSphs, while it is unlikely that it experienced any contribution from 
disrupted satellites like the more luminous dSphs with their absence of metal poor stars (see also Sect.~3.1). 

Finally, I note that, where ever old and intermediate age populations and/or a spread in metallicity is present in a dSph, 
population gradients have been detected in the sense of a central concentration of the 
metal rich, younger populations compared to the more extended old and metal poor component 
(Harbeck et al. 2001). 
The amplitude of this effect is, however, very different for individual galaxies: 
while for instance the Sculptor and Carina dSphs exhibit  clear radial separations of their stellar populations 
(Hurley-Keller et al. 1999; Harbeck et al. 2001; Tolstoy et al. 2004; Koch et al. 2006), 
other systems like Leo~II do not show any significant radial metallicity nor age gradients (Koch et al. 2007b).

\subsection{Chemical elements from high-resolution studies}
Given the faintness of the dSphs, these galaxies have long since evaded observations 
with high-resolution spectrographs, which are invaluable to perform detailed chemical abundance studies 
and to gain insight in the dominant modes and time scales of SF in the dSphs. 
This changed with the advent of larger telescopes of the 8-m class, and the first ground breaking 
study was that of Shetrone et al. (2001, 2003) who targeted 32 stars in seven of the nine luminous 
dSphs known at that time. Since then, the number of abundance data from high-resolution studies 
has vastly increased, and yet the information is sparse compared to the low-resolution metallicity 
measurements. As of today, chemical element ratios in approximately 100 red giants in {\em all}
of the nine luminous dSphs have been published in the literature (Shetrone et al. 2001, 2003, 2009; 
Sadakane et al. 2004; Geisler et al. 2004; Monaco et al. 2005; Koch et al. 2007d, 2008a). 
In the following plots we also include the data of Letarte (2007) for 82 stars in the  Fornax dSph. 
The situation for the ultrafaint galaxies is much sparser at present and only two stars in the faint 
Hercules dSph have been published to date in high-resolution mode (Koch et al. 2008b; Fig.~3),  although many ambitious programs are underway and the census is bound to increase quickly.

\subsubsection{The $\alpha$ elements}
The $\alpha$-elements (O,Mg,Si,Ca,Ti) are produced in core-collapse SNe of type II
that constitute the end phases, read: deaths,  of massive stars  above 8 M$_{\odot}$ 
 on negligible time scales. 
Iron can be formed in both SNe Ia and, in lesser amounts, in SNe II. 
SNe Ia are a consequence of mass transfer from giant companions on C/O white dwarfs so that 
one is dealing with a much lower mass regime and consequently longer lifetimes. 
Thus most of the iron at Solar metallicities originates from the long lived Ia,  while in metal poor stars it derives from SNe II,  
since the former were not present to enrich the early generations of stars, yet.

Literature comparisons 
often group together the light element abundances into a single [$<$Mg,Si,Ca,Ti$>$/Fe] 
ratio, which is perilous since the individual elements can be produced through different channels. 
In particular, Mg and O are formed during the hydrostatic C- and ensuing Ne-burning phases in the 
massive progenitors, while Si, Ca, and (presumably) Ti are formed explosively during the 
SN event itself 
(e.g., Timmes 1995; Woosley \& Weaver 1995; McWilliam 1997). 
As  pointed out by Shetrone et al. (2003), Venn et al. (2004; and reference therein), 
each of the $\alpha$-element to iron ratios follows slightly different trends; for instance the [Mg/Fe] ratio 
shows a broader, presumably real, cosmic scatter than the elements Ca or Ti (see also Koch et al. 2008a). 
In Fig.~5 I thus show the [Ca/Fe] ratio as representative of the $\alpha$-elements. 
\begin{figure}
\begin{center}
\epsfxsize=1\hsize\epsfbox{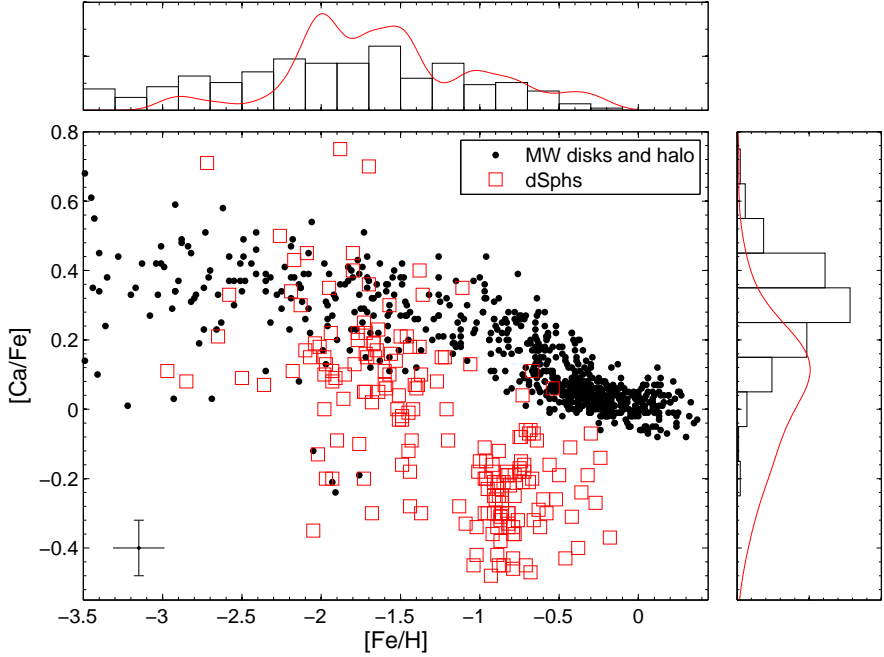}
\end{center}
\caption{$\alpha$-abundance ratios in Galactic stars (black dots) and dSph red giants (red squares). 
The dSph data are from Shetrone et al. 
(2001, 2003, 2009); Geisler et al. (2004);  Sadakane et al. (2004); Monaco et al. (2005); Letarte (2007); 
Koch et al. (2007d, 2008a,b), while Galactic stars were taken from Gratton \& Sneden (1988, 1994); 
Edvardsson (1993);  
McWilliam  et al. (1995);  Ryan et al. (1996); Nissen \& Schuster (1997);  McWilliam (1998);  
Hanson et al. (1998); Burris et al. (2000); Prochaska et al. (2000); Fulbright (2000, 2002); Stephens 
\& Boesgaard (2002); Johnson (2002); Bensby et al. (2003); Ivans et al. (2003); Reddy et al. (2003). 
Additional histograms show the halo distributions (black) in [Fe/H] and [Ca/Fe] in comparison to the generalized 
distributions (red lines) for the dSph stars, which have been weighted by measurement errors and by the 
number of targets per galaxy, such as not to introduce any observational bias.}
\end{figure}
Abundance studies carried out by different groups necessarily apply different techniques and 
input data, in particular regarding atomic parameters (such as $\log\,gf$ values vs. differential 
abundance studies that do not rely on these insecure values; Koch \& McWilliam 2008), model 
atmospheres (spherical vs. plane-parallel, Kurucz vs. MARCS) or the choice of atomic line lists.  
Throughout the following plots I did not attempt  to homogenize the abundance data from the literature accounting for these  
various approaches in the analyses. This will ultimately lead to an increased scatter 
among the dSph stars' abundances, of the order of the measurement uncertainties (Venn et al. 2004).  

The {\em Galactic} abundance distribution (black dots in Fig.~5) is well explicable in terms of 
a simple time delay model of chemical evolution (Tinsley 1979; Matteucci 2003; see also the review by McWilliam 1997). 
In this context, the enhanced value of [$\alpha$/Fe] $\sim  +0.4$ dex in the halo at low metallicities, widely dubbed the ``plateau'', 
is consistent with an early SF burst in the halo, 
which led to a high rate of early SNe II that produced a high amount of $\alpha$-elements 
and little iron. After a  delay of ca. 1 Gyr, the longer lived SNe Ia started contributing, thereby enriching 
the ISM and subsequent generations of stars with iron without producing $\alpha$-elements, which ultimately reflects in a decline of 
the [$\alpha$/Fe] abundance {\em ratio}. In the halo, this downturn occurs at [FeH]$\sim -1$  dex. 
Ultimately, this ratio is a delicate function of the initial mass function (IMF), the galaxies' SF histories, the 
involved SNe time scales,  as well as the time scale for mixing the SNe ejecta into the ISM (Matteucci (2003) 

One of the first and unprecedented notions of Shetrone et al. (2001) was that the 
[$\alpha$/Fe] abundance ratios in the dSph stars are systematically lower than those 
in Galactic halo stars of the same [Fe/H].   
The natural explanation for the low [$\alpha$/Fe] 
is the much lower SF rate in the dSphs (Unavane et al. 1996; Matteucci 2003; Lanfranchi \& Matteucci 2004; Sect.~3.1). 
Since SF was merely simmering in these galaxies, 
they had simply fewer SNe II to start with and thus much less  $\alpha$-elements 
were produced for the first generations of stars. By the time the SNe Ia started exploding, 
the ISM had not been enriched with iron from the SNe II to as high metallicities as in the halo. 
Consequently, the ``knee'' in the dSphs occurs at relatively low [Fe/H], e.g., at $\sim -1.8$ dex in Sculptor 
(Geisler et al. 2007) or $\sim -1.6$ in Carina (Fig.~6; Lanfranchi \& Matteucci 2006; Koch et al. 2007d,2008a). 
Alternatively, the low $\alpha$-abundances at intermediate metallicities are explicable through SF events 
with only low total masses involved. In events forming only a few 1000 M$_{\odot}$, very massive stars 
are unlikely to form at all (assuming a standard IMF). Since these stars are an efficient nucleosynthetic 
source of the $\alpha$-elements, their absence would lead to significantly lower [$\alpha$/Fe]
ratios compared to an environment with a high-mass SF and a fully sampled IMF like the halo 
(Woosley \& Weaver 1995; Shetrone et al. 2003; Koch et al. 2008a,b). Thus SF in, at least some, dSphs 
is likely to proceed on small scales.

The depletion of the dSphs'  [$\alpha$/Fe] ratios with respect to halo stars 
 was often cited as evidence that there was only little contribution of systems like 
the present-day dSphs to the Galactic halo build-up  at intermediate metallicities (which 
is strictly at odds with the observed current accretion of a dSph -- the Sagittarius dwarf; Ibata et al. 1994).  
In all such comparisons one should also strictly bear in mind that also the halo is distinct in its inner and outer components 
(Pritzl et al. 2005;  Carollo et al. 2007; Geisler et al. 2007; and references therein). 
This discrepancy is particularly pronounced at the metal rich end above [Fe/H]$\sim -1.5$ dex, where 
there is no overlap and the dSph stars exhibit [$\alpha$/Fe] ratios lower by up to 0.6 dex 
compared to the MW stars. 
On the other hand, Shetrone et al. (2003) reported that two of their dSph stars (in Leo~I and Sculptor) at 
$-1.5$ and $-2.0$ dex exhibit halo-like abundance ratios. 
With the increasing accumulation of larger data sets in the dSphs, stars were indeed found to partially  overlap with 
the Galactic halo population. 
As Fig.~5 indicates there are already a few dSph stars found with halo-like, that is enhanced, 
[$\alpha$/Fe] ratios at [Fe/H]=$-1.5$ dex and a handful that overlap below $-2$ dex; 
moreover, recent data (Frebel et al. 2009) indicate a significant overlap with the 
{\em metal-poor halo} below $-2$ dex and the dSph patterns appear to start resembling those of the halo around $-1.8$ dex. 
Thus it  appears likely that any contribution of stars from systems like the dSphs must have occurred at very early times (that is, 
at low metallicities). This is in concordance with the idea that the {\em ultrafaint}, more metal poor dSphs could have donated 
a fraction of the metal-poor Galactic halo (see Sect.~3.1, item {\em 4}), or at least that they had experienced similar 
formation and enrichment histories. 

There is a notable overlap with a few chemically  peculiar halo stars that stand out in terms of 
of a relatively strong depletion in the $\alpha$-elements, down to [$\alpha$/Fe]=$-0.2$ at 
[Fe/H]$\sim -2$ dex (Carney et al. 1997; Ivans et al. 2003). 
While the $\alpha$-elements in these stars are very similar to those found in a number of the dSphs (Fig.~5), 
not all of their heavy element patterns agree completely with those in the dSph stars 
(Sect.~3.2.2.; see also discussion in Ivans et al. 2003).  
The progressive detection of stars with strong element depletions (like LG04\_001826 in Carina with 
[Ca/Fe]=$-0.16$ and [Mg/Fe]=$-0.90$ at [Fe/H]=$-1.5$; Koch et al. 2008a) then reinforces the idea that 
complex dSphs are a possible source for the donation of chemically  peculiar stars to the halo. 
Moreover, few dSph stars exhibit  [$\alpha$/Fe] ratios well in excess of +0.4 dex,  
which is also found amongst the very metal poor halo stars below 
$\sim -3.5$  dex (e.g., McWilliam et al. 1995). 

An intriguing individual case is the Carina dSph that is {\em unique} among the Local Group dSphs in terms of 
its {\em episodic} SF history, in which 
periods of active SF are interrupted by extended quiescent phases of 
negligible SF activity (Smecker-Hane et al. 1994; Monelli et al. 2003; Tolstoy et al. 2003; 
Koch et al. 2006; see also Sect.~3.1). 
These periods have been photometrically well established 
and reflect, e.g., in the presence of multiple MSTOs representative of populations of approximately 0.6, 5, and 12  Gyr. 
Accordingly, its peculiarity also shows up in its chemical abundance ratios: this galaxy shows a very 
broad scatter in the [$\alpha$/Fe] abundance ratios at a given metallicity (Shetrone et al. 2003; Koch et al. 2007d,2008a), 
both with respect to the other 
dsph galaxies, and also in relation to model predictions of chemical evolution (Fig.~6; Gilmore \& Wyse 1991; 
Lanfranchi \& Matteucci 2004, 2006). 
\begin{figure}
\begin{center}
\epsfxsize=0.49\hsize\epsfbox{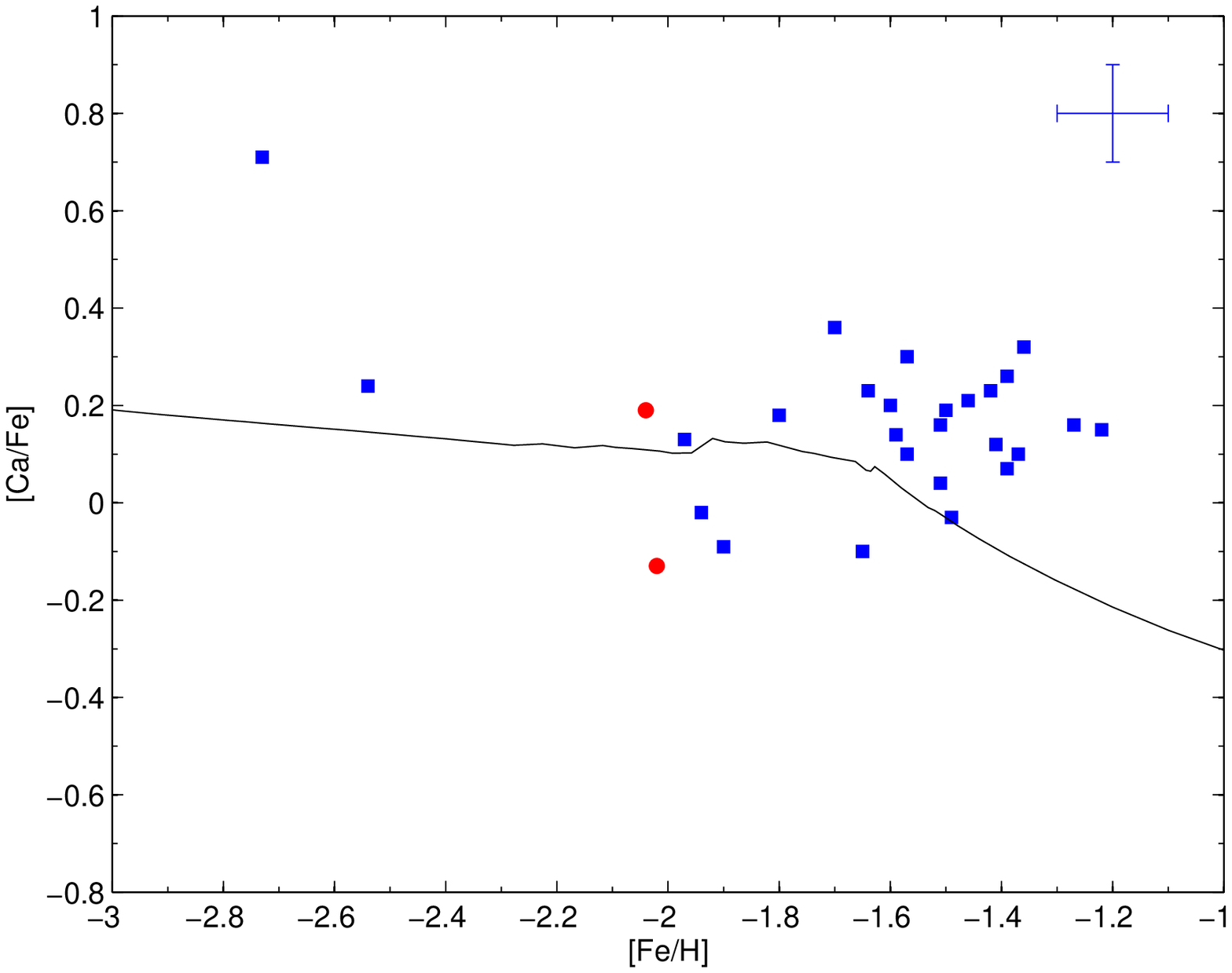}
\epsfxsize=0.49\hsize\epsfbox{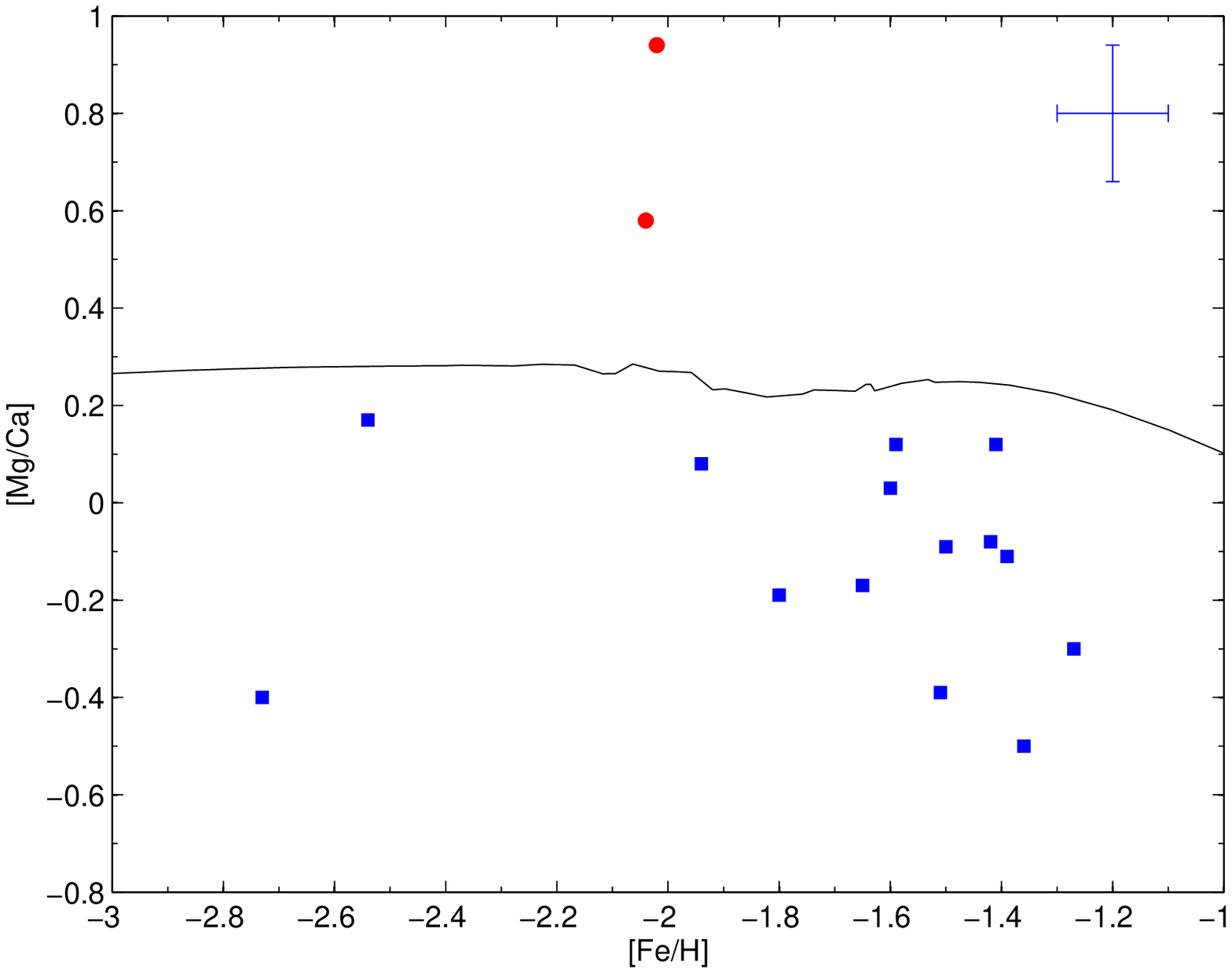}
\end{center}
\caption{[Ca/Fe] and [Mg/Ca] abundance ratios of red giants in  Carina (blue 
symbols; Shetrone et al. 2003; Koch et al. 2007d, 2008a) 
and Hercules (red symbols; Koch et al. 2008b). The solid lines are the model predictions  of Lanfranchi \& Matteucci (2006) 
for Carina. A typical error bar is indicated in the top right corners.}
\end{figure}

In the simple time-delay models, each SF bursts drives up the $\alpha$-element production, while 
Fe is continuously produced in the SNe Ia, which leads to an increase in the [$\alpha$/Fe]  ratio.  
Quiescent phases, however, do not have any SNe II and no $\alpha$-contribution, while Fe is still being produced in 
the Ia SNe.  The overlap of these episodes would then lead to a broad range in the abundance ratio over the whole 
metallicity range covered.  In Fig.~6 I show the model predictions by Lanfranchi \& Mateucci (2006) that are 
characterized by two SF bursts, efficient gas outflows (galactic winds; Sect.~3.1) and that were tailored 
to match Carina's evolution by using the observed MDF of Koch et al. (2006) and the chemical abundance data 
of Shetrone et al. (2003). Despite a good fit of  the overall MDF (Koch et al. 2006) and succeeding in the reproduction of 
an overall trend, the models do not represent the [Ca/Fe] abundance ratios in detail. In particular, the models predict the location of the downturn to 
occur at lower metallicities (by 0.2--0.3 dex) than is observed, leaving the impression of an underestimate of  
[Ca/Fe] at higher [Fe/H] with respect to the data points. The [Mg/Ca] ratio is in turn systematically overpredicted 
by approximately 0.3 dex. The exact times for onset, duration and cessation of the SF epochs as well as details 
of the SF and wind efficiencies are thus delicate governing parameters that have to account for as comprehensive 
abundance distributions as available and observable. 

Another attractive explanation for the large abundance spread in a dSph is the occurrence of stochastical SF 
on small scales. In cases, where every SF burst only converts small amounts of gas into stars
the IMF remains only incompletely and 
statistically sampled so that very massive stars may or may not be formed and thus be available as sources 
for the subsequent $\alpha$-element productions (e.g., Carigi et al. 2008).  
Such scenarios can be coupled with {\em spatially inhomogeneous} enrichment and a poor 
mixing of the SNe ejecta into the ISM (e.g., Marcolini et al. 2006), 
which then would reproduce dSph chemical abundance trends without the need to invoke 
galactic winds. 
Either enrichment scenario points to the complex chemical evolution of the dSphs. 
\subsubsection{Heavy elements}
Elements heavier than the iron peak are mainly produced through processes  
that  are distinguished by the rate of {\em neutron capture} relative to the time scale for 
the $\beta$-decay in the nuclei. In essence, the  $s$-process (for {\em s}low) takes place 
in environments with low neutron densities, such as low-mass (1--3 M$_{\odot}$) asymptotic giant 
branch (AGB)  stars (e.g., Gallino et al. 1998). Typical $s$-process elements are Y, Sr, Ba, La and Pb. 
The other main source for heavy element nucleosynthesis is the $r$-process ({\em r}apid) that occurs in 
environments dominated by high neutron fluxes such as the SNe II explosions or neutron star-neutron star 
mergers (see Qian \& Wasserburg 2007 for a recent review). 
An element often cited as the archetypical $r$-process element is Eu. 
One should keep in mind, however, that attributes such as Ba as an $s$-process element strictly hold for the Sun, while 
in the early universe, i.e., in metal-poor stars, all heavy elements can be expected to be produced in the $r$-process. 
At those early times, there were simply no longer-lived AGB stars present, yet, to contribute any $s$-process material. 
In this context, abundance studies of metal poor stars  and investigations in the dSphs are also invaluable to 
constrain possible nucleosynthetic production sites for the chemical elements (e.g., Thielemann et al. 2001; 
McWilliam et al. 2003; Frebel et al. 2009)
Individual [$s$/$r$] ratios such as [Ba/Eu] (Fig.~7) are then an important diagnostic for the relative interplay of 
the AGB enrichment on longer time scales and the fast SNe II enrichment and are also well suited to 
discern phases of more intense SF (with more SNe II) from quiescent phases (with AGB enrichment only). 
\begin{figure}
\begin{center}
\epsfxsize=1\hsize\epsfbox{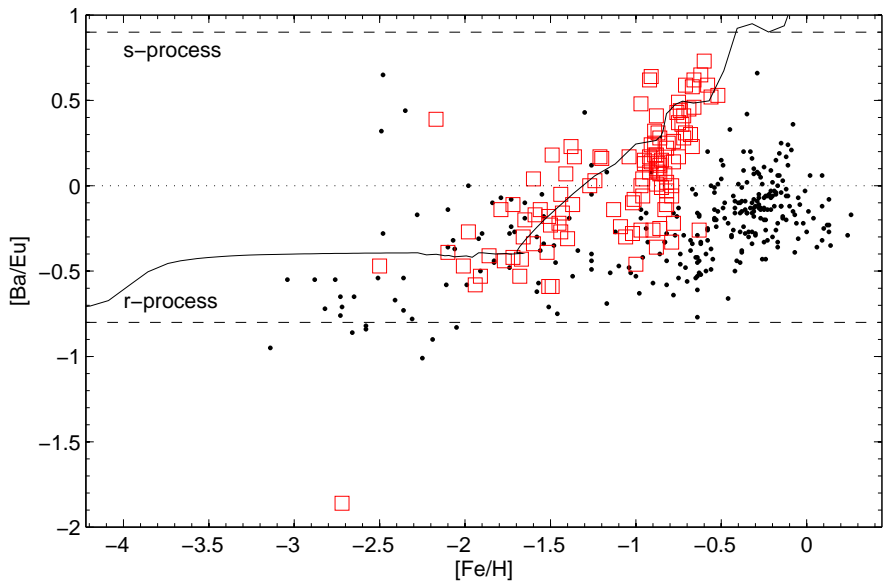}
\end{center}
\caption{[Ba/Eu] ratios using the same data as in Fig.~5. Shown as a solid line is the model for Carina by Lanfranchi et al.  (2008).}
\end{figure}

The trend seen in Fig.~7 confirms what was mentioned with regard to the [$\alpha$/Fe] ratio (Sect.~3.2.1). 
That is, dSphs are chemically distinct from the halo: while the halo rises from an r-process dominated 
environment towards s-process contributions at higher metallicities around $-$1, the $s$-process contribution 
in the dSphs rises earlier, at lower [Fe/H] of approximately $-1.7$ dex. 
This is due to the AGB stars contributing the $s$-process elements only after a longer delay, by which 
time the ISM was already enriched in Fe by the SNe II. 
Due to the low SF efficiency and thus less SNe~II in the dSphs, however, not much chemical enrichment had occurred 
by the time the AGBs started to contribute the s-process. 
The model prediction for Carina (solid line in Fig.~7) represents the moderately metal poor Carina well compared to the 
poor fit of the $\alpha$-data (Fig.~6). It is also worth noticing that Fornax experiences a much steeper upturn at much 
higher metallicities around $-1.2$ dex (in extrapolation).  

The addition of heavy element abundance information is crucial for a comparison with the Galactic halo. 
In order to plausibly argue, whether or not any component of the halo 
could have been donated by disrupted dSph systems, or shared  similar modes of SF, an overlap has to be present 
in essentially all chemical abundance patterns. 
In their study of chemically peculiar halo stars, Ivans et al. (2003) note $\alpha$, Cu and Zn abundances 
in the $\alpha$-depleted halo star BD+80$^{\circ}$~245 (Carney et al. 1997) that resemble those of a dSph
star in Carina (Shetrone et al. 2003), whereas there are significant differences of up to 2 dex in the $r$- and $s$-process 
abundance ratios. This and other examples given in Ivans et al.  (2003) efficiently rule out a connection 
of those halo stars with the dSphs. 
\subsubsection{Complex abundance ratios in complex dSphs}
As mentioned above, each individual galaxy exhibits intriguing and complex abundance patterns. 
Hence the statement that no two dSphs are alike and each has its own, special history still holds (Grebel 1997). 
One example is Carina in terms of its episodic SF and the resulting abundance scatter due to small-scale 
chemical evolution  in small associations. 
This galaxy also hosts one chemically peculiar star that is distinct in a strong depletion in all $\alpha$-elements. 
Also Dra~119, the most metal poor star in a luminous dSph to date, for which detailed abundance information is 
available ([Fe/H]=$-$2.95 dex; Shetrone et al. 2001; Fulbright et al. 2004) is very distinct in that it essentially  lacks all 
heavy elements beyond nickel. 

Albeit at higher metallicities, two stars in the Hercules dSph 
share similar properties: These stars are highly depleted in heavy elements, with [Ba/Fe]$<-2$ dex. 
Common to these stars is also that they exhibit high abundances of the hydrostatic $\alpha$-elements like O and Mg, while explosive elements 
like Ca are normal to deficient, leading to unusually high [Mg/Ca] ratios of +0.6--0.9 dex. 
A detailed interpretation of these pattern relies on the input nucleosynthetic models and is sensitive to stellar 
yields, the metallicity of the SNe progenitors and the treatment 
of rotation (e.g., Woosley \& Weaver 1995; Chiappini et al. 2003; Hirschi et al. 2005; 
Kobayashi et al. 2006; Heger \& Woosley 2008). 
The significance of this is that such high [Mg/Ca] ratios are understood by an enrichment of these stars through very massive 
progenitors, in the mass regime of $>20$ M$_{\odot}$ up to perhaps 50 M$_{\odot}$. 
In an environment in which a high mass of gas was converted into gas 
so that the IMF is fully sampled, it is feasible that such high mass stars form. 
How then can the ISM retain the {\em chemically 
peculiar} pattern that is imprinted in the anomalous stars we observe today, e.g., in Hercules? 
This is only possible, again, if SF proceeded highly stochastically, 
with one or two very high mass stars forming (the observed patterns are in fact explicable with less than 3 of these SNe events; 
Koch et al. 2008b), while the lower mass SNe II might not have occurred. This is most likely to happen in very low-mass 
SF events (see also Carigi et al. 2008) in the regime of less than a few thousand M$_{\odot}$ in total stellar mass. 
Although it is statistically very unlikely that any such very high-mass stars are formed at all under a standard IMF 
(e.g., Miller \& Scalo 1979; see also  Shetrone et al. 2003; Koch et al. 2008a), 
{\em if} they happen to form,  early in the galaxy's history, they will effectively govern the enrichment patterns 
of the entire low-mass environment, which is the most likely scenario for Hercules' evolution to date. 
This is likely coupled with highly inhomogeneous spatial enrichment of the SNe~Ia and II;  the few massive SNe
explosions would then only  influence the immediate surrounding ISM out of which the subsequent stars are born 
(see also Marcolini et al. 2006,2008). 

Fig.~8  shows [Mg/Ca] ratios in the sample of Galactic and dSph stars introduced in Fig.~5. 
\begin{figure*}
\begin{center}
\epsfxsize=0.49\hsize \epsfbox{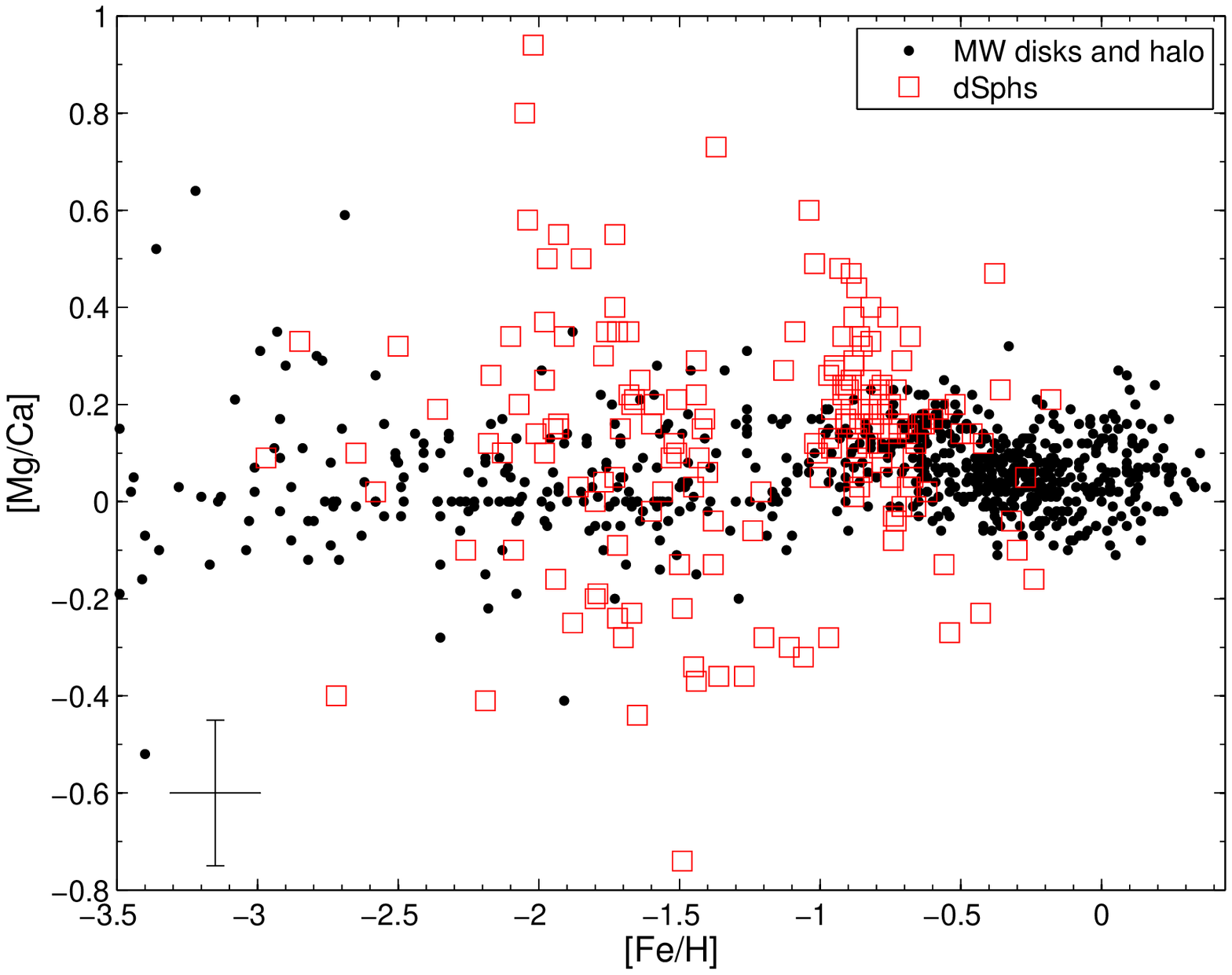}
\epsfxsize=0.49\hsize \epsfbox{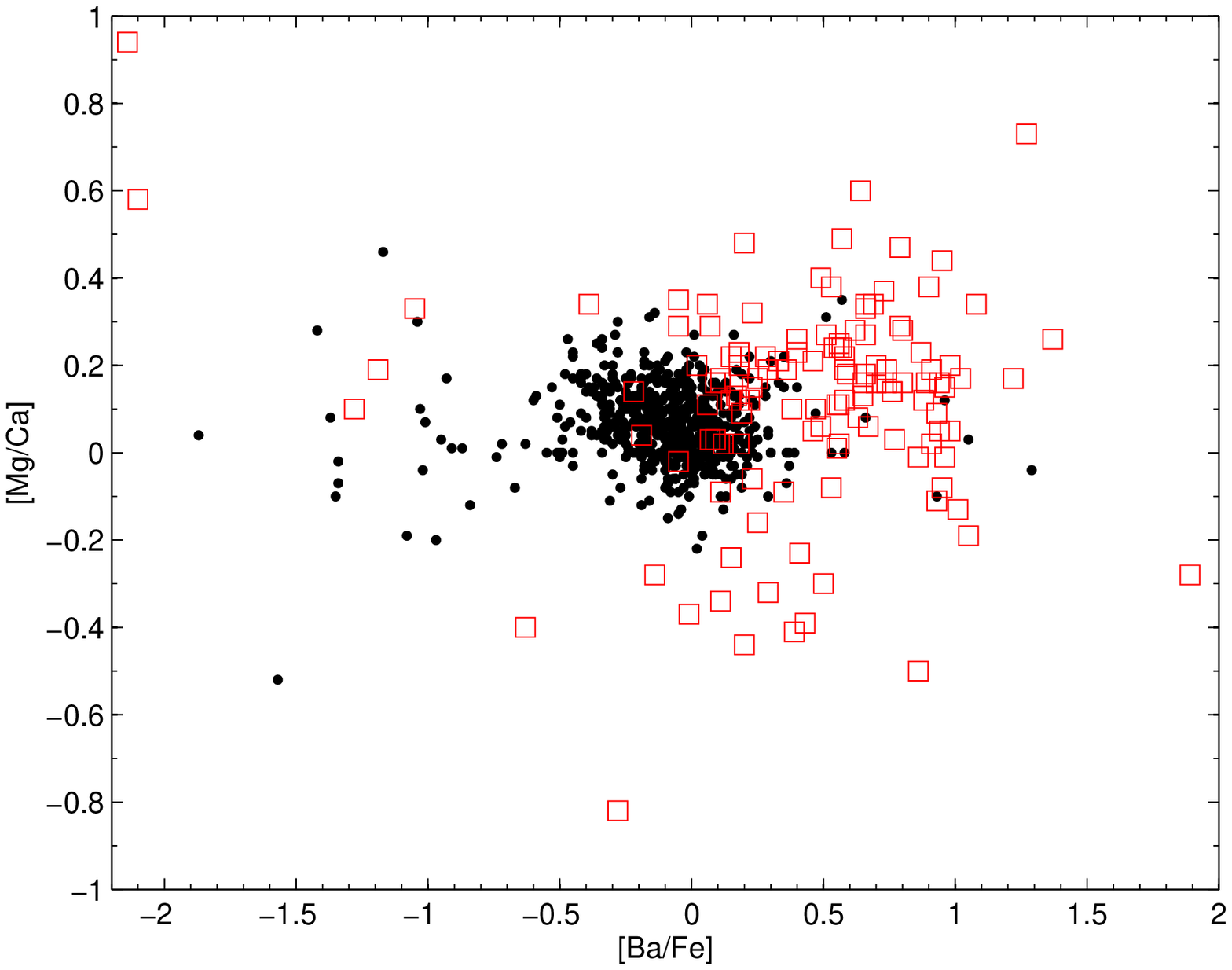}
\end{center}
\caption{[Mg/Ca]  abundance ratios as function of iron abundance [Fe/Fe] (left panel) 
and the heavy-element abundance ratio [Ba/Fe] (right panel). 
Shown are the  Galactic and dSph stars using the same symbols and data as in Fig.~5.  
In these plots,  chemically peculiar stars stand out by strong enhancements in the [Mg/Ca] ratio.}
\end{figure*}
At any given metallicity, there is a broad scatter present in this ratio in the dSph stars, although 
the generally large error bars on this ratio in the faint dSphs should be noted (indicated in Fig.~8). 
According to the yields of Heger \& Woosley (2008), SNe II with progenitor masses above 20 (25) M$_{\odot}$ 
can already produce  [Mg/Ca] ratios in excess of 0.3 (0.5) dex. 
Apart from the two Hercules stars and Dra~119, also  a handful of giants in the Leo II dSph 
between an [Fe/H] of $-1.7$ to $-2$ dex (Shetrone et al. 2009)  exhibit enhanced [Mg/Ca] ratios. 
It is interesting to note that Fornax contains a high fraction (ca. 1/5th of the sample) of [Mg/Ca]-enhanced stars above the Galactic, Solar value. 
It is thus conceivable that this galaxy experienced a higher number of massive SNe II than other dwarfs, which is feasible 
since Fornax is the most luminous dSph and it is governed by an extended SF history. 
On the other hand, the Hercules stars of Koch et al. (2008b) and 
Dra~119 (Fulbright et al. 2004) are significantly depleted in the chemical  elements heavier than Ni 
so that an additional conclusion is that very massive SNe II progenitors likely do not synthesize such heavy elements 
(see also Frebel et al. 2009). 
If the nucleosynthesis governing all of these galaxies was thus comparable, such depletions should be omnipresent in all 
strongly [Mg/Ca]-enhanced stars. 
As Fig.~8 (right panel) shows, there is no significant trend discernible -- most higher [Mg/Ca] stars show
enhancements in [Ba/Fe] that are typical of their metallicity. 
Thus one is facing two entirely different mechanisms: the stochastical enrichment with very massive star contributions 
on the one hand;  and on the other hand the integrated build-up of heavy elements coupled with the massive star 
contribution as expected from a fully sampled IMF in a massive system with continuous SF activity like Fornax. 

Another relevant pattern is for instance the [Co/Cr] abundance ratio, which is notably high in the Hercules stars (Koch et 
al. 2008b) 
such as usually found only in very metal poor Galactic halo stars, below [Fe/H]$\la$$-3$ dex (e.g., McWilliam et al. 1995).  
This can be interpreted as a signature of the aforementioned massive SNe~II enrichment, coupled with contributions from the 
hypothesized first, metal-free massive Population III stars (e.g., Beasley et al. 2003; Bromm \& Larson 2004; Frebel et al. 2007; 
Cohen et al. 2007). 
Thus the (ultra-) faint dSphs are in fact an attractive environment to search for signatures of the Population~III stars -- galaxies like 
Hercules could well show such evidence of primordial material and be the sites of the {\em first stars} in the universe. 
\section{Summary}
How similar is the newly discovered generation of ultrafaint galaxies to the classical, more luminous dSphs?
Are they mere extensions of the known, extreme properties of the latter? 
Their general characteristics as ``old, metal poor and dark matter dominated'' systems is in fact reminiscent 
of the traditional dwarfs, 
yet the ultrafaint galaxies are characterized by much lower luminosities, they are found to be 
more metal poor on average with broad abundance spreads, 
and have higher M/L ratios than the traditional satellites, thus appear to 
be even more strongly dark matter dominated. 

With the gathering of detailed chemical abundance data in the past few years, the classical picture of the 
dSphs as relatively simple, chemically homogeneous and well-mixed systems has drastically changed. 
Individual abundance patterns and the broad abundance scatter in several of these galaxies indicate that,  
{\em already on small scales},  there is room for local inhomogeneities through incomplete mixing processes 
of the SNe ejecta or an incomplete sampling of the IMF due to low-mass SF events. 
The dominance of such inhomogeneities was already found in many, more massive, dwarf irregular galaxies  
(e.g., Sextans B, Kniazev et al. 2005; or the SMC, Glatt et al. 2008).  
These patterns then underscore the complexities in the SF, enrichment and evolutionary processes on 
small scales in the dSphs. 
Moreover, in the ultrafaint galaxies there appears to exist a mode of SF that has not been identified in any other of the dSphs 
before (though visible, e.g., in the Galactic halo), that is, we may be able to  see here the potential signatures of the first stars, 
imprinted in the abundance anomalies of the subsequent generations of stars observed today. These galaxies thus hold the keys 
for tracing the fossil records of SF. 

Furthermore, the growing body of chemical evidence confirms the early findings of Shetrone et al. (2001, 2003) and Fulbright  (2002)  
that  the discrepancies in the abundance ratios between the dSps and the Galactic halo persist {\em at intermediate 
metallicities}. 
However, from more recent data the earlier surmised overlap with the Galactic halo {\em at the metal poor end} emerges, 
thereby reinforcing the importance of dSph-like systems in the build-up of {\em parts of the halo} (see also Bell et al. 2008).  
This is then also consistent with the very metal poor character of the {\em ultrafaint} satellites.  
Yet these systems clearly cannot account for a full build-up of the stellar halo, which would  rather require few (1--2) early, 
massive (LMC type) accretions, while the present-day (ultrafaint) dSphs are the mere survivors of ancient accretion 
processes  (see also Zinn \& West 1978; Unavane et al. 1996; Bullock \& Johnston 2005; Robertson et al. 2005). 

Where do we go from here? 
In the near future, the MW dSph satellite family will most certainly vastly increase, opening intriguing new windows to 
study the detailed, complex properties of these cosmological substructures on small scales. 
Such new discoveries will be characterized by progressively faint magnitudes so that straightforward spectroscopic 
follow-up observations are very time-expensive. 
In this context it is timely to acknowledge the great powers and availability of state-of-the-art low- and high-resolution 
instruments at our generation's 8-10m class telescopes: amongst these, the FLAMES/UVES multi-object spectrograph 
at the ESO/VLT, the MIKE and MOE Echelle spectrographs at Magellan, as well as the DEIMOS and HIRES instruments 
at Keck have delivered the vast amounts of invaluable data I presented in this review.  
And yet it should be mentioned that, e.g., the first high-resolution spectroscopic study in brighter giants of  a faint dSph, viz.  
Hercules (Fig.~3; Sect.~3.2.3) already required 4--6 hours integration at a 6.5m telescope to obtain 
sufficient signal-to-noise ratios (Koch et al. 2008b). 
Deriving detailed abundance ratios a large number of even fainter stars in the ultrafaint dSphs then 
calls for dedicated programs and future generations of powerful, preferentially multi-object, facilities 
at telescopes of the ELT, TMT and GMT class. 

\subsection*{Acknowledgements}
First of all I would like to thank the Astronomische Gesellschaft for granting me the honor 
of the Ludwig Biermann award. I gratefully acknowledge support from and invaluable discussions with 
Eva K. Grebel, Andrew McWilliam, Mark Wilkinson, Gerry Gilmore, Rosie Wyse, R. Michael Rich, Francesca Matteucci, 
Andrea Marcolini, Nicolas Martin and Gustavo Lanfranchi.  
Finally I would like to thank the Swiss National Science Foundation, SNF,  
which funded much of my own work presented in this review.
 
\subsection*{References}

{\small
\bref
Aaronson, M.\ 1983, ApJL, 266, L11 

\bref
Armandroff, T.~E., \& Da Costa, G.~S.\ 1991, AJ, 101, 1329 

\bref
Armandroff, T.~E., \& Zinn, R.\ 1988, AJ, 96, 92 

\bref
Battaglia, G., Helmi, A., Tolstoy, E., Irwin, M., Hill, V., \& Jablonka, P.\ 2008a, ApJL, 681, L13 

\bref
Battaglia, G., Irwin, M., Tolstoy, E., Hill, V., Helmi, A., Letarte, B., \& Jablonka, P.\ 2008b, MNRAS, 383, 183

\bref
Beasley, M.~A., Kawata,  D., Pearce, F.~R., Forbes, D.~A., \& Gibson, B.~K.\ 2003, ApJL, 596, L187 

\bref
Beers, T.~C., \& Christlieb, N.\ 2005, ARA\&A, 43, 531 

\bref
Bell, E.~F., et al.\ 2008, ApJ, 680, 295 

\bref
Bellazzini, M., Ferraro, F.~R., Origlia, L., Pancino, E., Monaco, L., \& Oliva, E.\ 2002, AJ, 124, 3222 

\bref
Bellazzini, M., et al.\ 2008, AJ, 136, 1147 

\bref
 Belokurov, V., et   al.\ 2006a, ApJL 647, L111 

\bref
 Belokurov, V., et   al.\ 2006b, ApJL 642, L137 

\bref
Belokurov, V., Evans,  N.~W., Irwin, M.~J., Hewett, P.~C.,  \& Wilkinson, M.~I.\ 2006c, ApJL, 637, L29 

\bref
Belokurov, V., et al.\ 2007a, ApJL, 654, 897 

\bref
Belokurov, V., et al.\ 2007b, ApJ, 658, 337 

\bref
Belokurov, V., et al.\ 2008, ApJL, 686, L83

\bref
Bensby, T., Feltzing, S., \& Lundstr{\"o}m, I.\ 2003, A\&A, 410, 527 

\bref
Binney, J., \& Tremaine, S.\ 1987, Princeton, NJ, Princeton University Press, 1987, 747 p.,  

\bref
Bosler, T.~L., Smecker-Hane, T.~A., \& Stetson, P.~B.\ 2007, MNRAS, 378, 318 

\bref
Bromm, V., \& Larson, R.~B.\ 2004, ARA\&A, 42, 79 

\bref
Bullock, J.~S., \& Johnston, K.~V.\ 2005, ApJ, 635, 931 

\bref
Burris, D.~L., Pilachowski, C.~A., Armandroff, T.~E., Sneden, C., Cowan, J.~J., \& Roe, H.\ 2000, ApJ, 
544, 302 

\bref
Cannon, R.~D., Hawarden, T.~G., \& Tritton, S.~B.\ 1977, MNRAS, 180, 81P 

\bref
Carigi, L., Hernandez, X., \& Gilmore, G.\ 2002, MNRAS, 334, 117 

\bref
Carigi, L., \& Hernandez, X.\ 2008, MNRAS, 390, 582 

\bref
Carney, B.~W., Wright, J.~S., Sneden, C., Laird, J.~B., Aguilar, L.~A., 
\& Latham, D.~W.\ 1997, AJ, 114, 363 

\bref
Carollo, D., et al.\  2007, Nature, 450, 1020

\bref
Carrera, R., Gallart, C., Pancino, E., \& Zinn, R.\ 2007, AJ, 134, 1298 

\bref
Carretta, E., \& Gratton, R. 1997, A\&AS, 121, 95

\bref
Chiappini, C., Matteucci, F., \& Meynet, G.\ 2003, A\&A	, 410, 257 

\bref
Cohen, J.~G., McWilliam, A., Christlieb, N., Shectman, S., Thompson, I., Melendez, J., Wisotzki, L., \& Reimers, D.\ 2007, ApJL, 
 659, L161 

\bref
Cohen, J.~G., Christlieb, 
N., McWilliam, A., Shectman, S., Thompson, I., Melendez, J., Wisotzki, L., 
\& Reimers, D.\ 2008, ApJ, 672, 320 

\bref
Cole, A.~A., Smecker-Hane, T.~A., \& Gallagher, J.~S., III 2000, AJ, 120, 1808 

\bref
Cole, A.~A., Smecker-Hane,  T.~A., Tolstoy, E., Bosler, T.~L., \& Gallagher, J.~S.\ 2004, MNRAS, 347, 367 

\bref
Coleman, M.~G., et al.\  2007, ApJL, 668, L43 

\bref
C{\^o}t{\'e}, P., Oke, J.~B., \& Cohen, J.~G.\ 1999, AJ, 118, 1645 

\bref
de Jong, J.~T.~A., Rix, H.-W., Martin, N.~F., Zucker, D.~B., Dolphin, A.~E., Bell, E.~F., 
Belokurov, V., \& Evans, N.~W.\ 2008, AJ, 135, 1361 

\bref
Dekel, A., \& Silk, J.\ 1986, ApJ, 303, 39 

\bref
Dekel, A., \& Woo, J.\ 2003, MNRAS, 344, 1131 

\bref
Dotter, A., Chaboyer, B., Jevremovi{\'c}, D., Kostov, V., Baron, E., \& Ferguson, J.~W.\ 2008, ApJS, 
178, 89 

\bref
Edvardsson, B., Andersen, J., Gustafsson, B., Lambert, D.~L., Nissen, P.~E., \& Tomkin, J.\ 1993, A\&A, 
275, 101 

\bref
Faria, D., Feltzing, S., Lundstr{\"o}m, I., Gilmore, G., Wahlgren, G.~M., Ardeberg, A., \& Linde, P.\ 2007, A\&A, 465, 357 

\bref
Fellhauer, M., et al.\ 2007, MNRAS, 375, 1171 

\bref
Ferguson, A.~M.~N., Gallagher, J.~S., \& Wyse, R.~F.~G.\ 2000, AJ, 120, 821 

\bref
Font, A.~S., Johnston, K.~V., Bullock, J.~S., \& Robertson, B.~E.\ 2006, ApJ, 638, 585 

\bref
Frebel, A., Johnson,  J.~L., \& Bromm, V.\ 2007, MNRAS, 380, L40 

\bref 
Frebel, A., Simon, J.~D., Geha, M., \& Willman, B.\ 2009, ApJ, submitted (arXiv:0902.2395)

\bref
Fulbright, J.~P.\ 2000, AJ, 120, 1841 

\bref
Fulbright, J.~P.\ 2002, AJ, 123, 404 

\bref
Fulbright, J.~P.,  Rich, R.~M., \& Castro, S.\ 2004, ApJ, 612, 447 

\bref
Gallagher, J.~S., Madsen, G.~J., Reynolds, R.~J., Grebel, E.~K., \& Smecker-Hane, T.~A.\ 2003, ApJ, 588, 326 

\bref
Gallart, C., Freedman, W.~L., Aparicio, A., Bertelli, G., \& Chiosi, C.\ 1999, AJ, 118, 2245 

\bref
Gallino, R., Arlandini, C., Busso, M., Lugaro, M., Travaglio, C., Straniero, O., Chieffi, A., \& Limongi, M.\ 1998, ApJ, 497, 388 

\bref
Geisler, D., Smith, V.~V., Wallerstein, G., Gonzalez, G., \& Charbonnel, C. 2005, AJ, 129, 1428

\bref 
 Gilmore, G., Wilkinson, M.~I., Wyse, R.~F.~G., Kleyna, J.~T., Koch, A., Evans, N.~W., 
\& Grebel, E.~K.\ 2007, ApJ, 663, 948 

\bref
Glatt, K., et al.\ 2008, AJ, 136, 1703 

\bref
Goerdt, T., Moore, B., Read, J.~I., Stadel, J., \& Zemp, M.\ 2006, MNRAS, 368, 1073 

\bref
Gratton, R.~G., \& Sneden, C.\ 1988, A\&A, 204, 193 

\bref
Gratton, R.~G., \& Sneden, C.\ 1994, A\&A, 287, 927 

\bref
Grebel, E.~K.\ 1997, Reviews in Modern Astronomy, 10, 29 

\bref
Grebel, E.~K., Gallagher, J.~S., III, \& Harbeck, D.\ 2003, AJ, 125, 1926

\bref
Grebel, E.~K., \&  Gallagher, J.~S., III 2004, ApJL, 610, L89 

\bref
Grillmair, C.~J. 2008, ApJ, in press (astro-ph/0811.3965v1) 

\bref
Harbeck, D., et al.\ 2001, AJ, 122, 3092

\bref
Hanson, R.~B., Sneden, C., Kraft, R.~P., \& Fulbright, J.\ 1998, AJ, 116, 1286 

\bref
Heger, A., \& Woosley, S.~E.\ 2008, ApJS, submitted (arXiv:0803.3161)

\bref
Helmi, A., et al.\ 2006, ApJL, 651, L121 

\bref
Hensler, G., Theis, C., \& Gallagher,, J.~S., III.\ 2004, A\&A, 426, 25 

\bref
Hernquist, L.\ 1990, ApJ, 356, 359 

\bref
Hirschi, R., Meynet, G., \& Maeder, A.\ 2005, A\&A, 433, 1013 

\bref
Hurley-Keller, D., Mateo, M., \& Grebel, E.~K.\ 1999, ApJL, 523, L25 

\bref
Ibata, R.~A., Gilmore, G., \& Irwin, M.~J.\ 1994, Nature, 370, 194 

\bref
Ibata, R., Irwin, M., Lewis, G., Ferguson, A.~M.~N., \& Tanvir, N.\ 2001, Nature, 412, 49 

\bref
Ibata, R., Chapman, S., Irwin, M., Lewis, G., \& Martin, N.\ 2006, MNRAS, 373, L70 

\bref
Ibata, R., Martin, N.~F., Irwin, M., Chapman, S., Ferguson, A.~M.~N., Lewis, G.~F., 
\& McConnachie, A.~W.\ 2007, ApJ, 671, 1591 

\bref
Ikuta, C., \& Arimoto, N.\ 2002, A\&A, 391, 55 

\bref
Illingworth, G.\ 1976, ApJ, 204, 73 

\bref
Inman, R.~T., \& Carney, B.~W.\ 1987, AJ, 93, 1166 

\bref Ivans, I.~I., Sneden, C., James, C.~R., Preston, G.~W., Fulbright, J.~P., H{\"o}flich, P.~A., Carney, 
B.~W., \& Wheeler, J.~C.\ 2003, ApJ, 592, 906 

\bref
Johnson, J.~A.\ 2002, ApJS, 139, 219 

\bref
Kazantzidis, S., Mayer, L., Mastropietro, C., Diemand, J., Stadel, J., \& Moore, B.\ 2004, ApJ, 608, 663 

\bref
King, I.~R.\ 1966, AJ, 71, 64 

\bref 
Kirby, E.~N., Simon, J.~D., Geha, M., Guhathakurta, P., \& Frebel, A.\ 2008a, ApJL, 685, L43 

\bref
Kirby, E.~N., Guhathakurta, P., \& Sneden, C.\ 2008b, ApJ, 682, 1217 

\bref
Klessen, R.~S., Grebel, E.~K., \& Harbeck, D.\ 2003, ApJ, 589, 798 

\bref 
Kleyna, J., Wilkinson, M.~I., Evans, N.~W., Gilmore, G., \& Frayn, C.\ 2002, MNRAS, 330, 792 

\bref
Kleyna, J.~T., Wilkinson, M.~I., Gilmore, G., \& Evans, N.~W. 2003, ApJ, 588, L21

\bref
Kniazev, A.~Y., Grebel, E.~K., Pustilnik, S.~A., Pramskij, A.~G., \& Zucker, D.~B.\ 2005, AJ, 130, 1558 

\bref
Kobayashi, C., Umeda, H., Nomoto, K., Tominaga, N., \& Ohkubo, T.\ 2006, ApJ, 653, 1145 

\bref
Koch, A., Grebel, E.~K., Wyse, R.~F.~G., Kleyna, J.~T., Wilkinson, M.~I., 
Harbeck, D.~R., Gilmore, G.~F., \& Evans, N.~W. 2006, AJ, 131, 895

\bref
 Koch, A., Wilkinson, M.~I., Kleyna, J.~T., Gilmore, G.~F., Grebel, E.~K., Mackey, A.~D., Evans, 
N.~W., \& Wyse, R.~F.~G.\ 2007a, ApJ, 657, 241  

\bref
Koch, A., Grebel, E.~K., Kleyna, J.~T., Wilkinson, M.~I., Harbeck, D.~R., Wyse, R.~F.~G., \& Evans, 
N.~W., 2007b, AJ, 133, 270

\bref
Koch, A., Kleyna, J.~T., Wilkinson, M.~I., Grebel, E.~K., Gilmore, G.~F., Evans, N.~W., Wyse, 
R.~F.~G., \& Harbeck, D.~R.\ 2007c, AJ,  134, 566   

\bref
Koch, A., et al.\ 2007d, AN, 328, 652 

\bref
Koch, A., \& McWilliam, A.\ 2008, AJ, 135, 1551 

\bref
Koch, A., Grebel, E.~K., Gilmore, G.~F., Wyse, R.~F.~G., Kleyna, J.~T., Harbeck, D.~R., Wilkinson, 
M.~I., \& Wyn Evans, N.\ 2008a, AJ, 135, 1580 

\bref
Koch, A., McWilliam, A., Grebel, E.~K., Zucker, D.~B., \& Belokurov, V.\ 2008b, ApJL, 688, L13 

\bref
 Koch, A., et al.\ 2008c, ApJ, 689, 958 

\bref
Koch, A. et al. 2009, ApJ, 690, 453 

\bref
Koposov, S., et al.\  2007, ApJ, 669, 337 

\bref
Koposov, S., et al.\  2008, ApJ, 686, 279 

\bref
Kraft, R.~P., \& Ivans, I.~I.\ 2003, PASP, 115, 143 

\bref
Kroupa, P.\ 1997, New Astronomy, 2, 139 

\bref
Lanfranchi, G.~A., \& Matteucci, F.\ 2004, MNRAS, 351, 1338 

\bref
Lanfranchi, G.~A.,  Matteucci, F., \& Cescutti, G. \ 2006, A\&A, 453, L67 

\bref
Lanfranchi, G.~A., \& Matteucci, F.\ 2007, A\&A, 468, 927 

\bref
Lanfranchi, G.~A., Matteucci, F., \& Cescutti, G.\ 2008, A\&A, 481, 635 

\bref
Larson, R.~B.\ 1998, MNRAS, 301, 569 

\bref
Letarte, B. 2007, PhD Thesis, Rijksuniversiteit Groningen

\bref
Liu, C., Hu, J., Newberg, H., \& Zhao, Y.\ 2008, A\&A, 477, 139 

\bref
{\L}okas, E.~L.\ 2001, MNRAS, 327, L21 

 \bref
{\L}okas, E.~L.\ 2002, MNRAS, 333, 697 

\bref
{\L}okas, E.~L.\ 2009, MNRAS, in press (arXiv:0901.0715) 

\bref
Mac Low, M.-M., \& Ferrara, A.\ 1999, ApJ, 513, 142 

\bref
Majewski, S.~R.,  Ostheimer, J.~C., Kunkel, W.~E., \& Patterson, R.~J.\ 2000, AJ, 120, 2550 

\bref
Majewski, S.~R., et al.\ 2007, ApJL, 670, L9 

\bref
Marcolini, A., D'Ercole, A., Brighenti, F., \& Recchi, S.\ 2006, MNRAS, 371, 643 

\bref
Marcolini, A., D'Ercole, A., Battaglia, G., \& Gibson, B.~K.\ 2008, MNRAS, 386, 2173 

\bref
Martin, N.~F., Ibata, R.~A., Irwin, M.~J., Chapman, S., Lewis, G.~F., Ferguson, A.~M.~N., Tanvir, 
N., \& McConnachie, A.~W.\ 2006, MNRAS, 371, 1983 

\bref
Martin, N.~F., Ibata, R.~A., Chapman, S.~C., Irwin, M., \& Lewis, G.~F.\ 2007, MNRAS, 380, 281 

\bref 
Martin, N.~F., de Jong, J.~T.~A., \& Rix, H.-W.\ 2008, ApJ, 684, 1075 

\bref
Mateo, M., Olszewski, E.~W., Pryor, C., Welch, D.~L., \& Fischer, P.\ 1993, AJ, 105, 510 

\bref
Mateo, M.~L.\ 1998, ARA\&A, 36, 435 

\bref
Matteucci, F.\ 2003, ApS\&S, 284, 539 

\bref
McConnachie, A.~W., \& Irwin, M.~J.\ 2006, MNRAS, 365, 1263 

\bref
McConnachie, A.~W.,  et al.\ 2008, ApJ, 688, 1009 

\bref
McWilliam, A., Preston, G.~W., Sneden, C. \& Searle, L. 1995,  AJ, 109, 275

\bref
McWilliam, A.\ 1997, ARA\&A, 35, 503 

\bref
McWilliam, A.\ 1998, AJ, 115, 1640 

\bref
McWilliam, A., Rich, R.~M., \& Smecker-Hane, T.~A.\ 2003, ApJL, 592, L21 

\bref
Miller, G.~E., \& Scalo, J.~M.\ 1979, ApJS, 41, 513 

\bref
Monaco, L., Bellazzini, M., Bonifacio, P., Ferraro, F.~R., Marconi, G., Pancino, E., Sbordone, L., \& 
Zaggia, S. 2005, A\&A, 441, 141

\bref
Monelli, M., et al.\  2003, AJ, 126, 218 

\bref
Moore, B., Ghigna, S., Governato, F., Lake, G., Quinn, T., Stadel, J., \& Tozzi, P.\ 1999, ApJ, 524, L19 

\bref
Morrison, H.~L., Harding, P., Hurley-Keller, D., \& Jacoby, G.\ 2003, ApJL, 596, L183 

\bref
Mu{\~n}oz, R.~R., et  al.\ 2006a, ApJ, 649, 201 

\bref
Mu{\~n}oz, R.~R., Carlin, J.~L., Frinchaboy, P.~M., Nidever, D.~L., Majewski, S.~R., 
\& Patterson, R.~J.\ 2006b, ApJL, 650, L51 

\bref
Mu{\~n}oz, R.~R., Majewski, S.~R., \& Johnston, K.~V.\ 2008, ApJ, 679, 346 

\bref
Navarro, J.~F., Frenk, C.~S., \& White, S.~D.~M.\ 1997, ApJ, 490, 493 

\bref
Nissen, P.~E., \& Schuster, W.~J.\ 1997, A\&A, 326, 751 

\bref
Pe{\~n}arrubia, J., Navarro, J.~F., \& McConnachie, A.~W.\ 2008a, ApJ, 673, 226 

\bref
Pe\~narrubia, J., Navarro, J.~F., McConnachie, A.~W., \& Martin, N.~F.\ 2008b, ApJ, submitted (arXiv:0811.1579) 

\bref
Pont, F., Zinn, R., Gallart, C., Hardy, E., \& Winnick, R.\ 2004, AJ, 127, 840 

\bref
Prochaska, J.~X., Naumov, S.~O., Carney, B.~W., McWilliam, A., \& Wolfe, A.~M.\ 2000, AJ, 120, 2513 

\bref
Qian, Y.-Z., \& Wasserburg, G.~J.\ 2007, PhR, 442, 237 

\bref
Read, J.~I., \& Gilmore, G.\ 2005, MNRAS, 356, 107 

\bref
Read, J.~I., Wilkinson, M.~I., Evans, N.~W., Gilmore, G., \& Kleyna, J.~T.\ 2006, MNRAS, 367, 387 

\bref
Reddy, B.~E., Tomkin, J., Lambert, D.~L., \& Allende Prieto, C.\ 2003, MNRAS, 340, 304 

\bref
Robertson, B., Bullock, J.~S., Font, A.~S., Johnston, K.~V., 
\& Hernquist, L.\ 2005, ApJ, 632, 872 

\bref
Rosenberg, A., Saviane, I., Piotto, G., Aparicio, A., \& Zaggia, S.~R.\ 1998, AJ, 115, 648 

\bref
Ryan, S.~G., Norris, J.~E., \& Beers, T.~C.\ 1996, ApJ, 471, 254 

\bref
Sadakane, K., Arimoto, N., Ikuta, C., Aoki, W., Jablonka, P., \& Tajitsu, A. 2004, PASJ, 56, 1041

\bref
Salvadori, S., \& Ferrara, A. 2008, MNRAS, in press (astro-ph/0812.2151v1)

\bref
S{\'a}nchez-Salcedo, F.~J., Reyes-Iturbide, J., \& Hernandez, X.\ 2006, MNRAS, 370, 1829 

\bref
Schneider, R., Ferrara, A., Natarajan, P., \& Omukai, K.\ 2002, ApJ, 571, 30 

\bref
Searle, L., \& Zinn, R.\ 1978, ApJ, 225, 357 

\bref
Shetrone, M.~D., C\^ot\'e, P., \& Sargent, W.~L.~W.\ 2001, ApJ, 548, 592

\bref
Shetrone, M.~D., Venn, K.~A., Tolstoy, E., Primas, F.,  Hill, V., \& Kaufer, A. 2003, AJ, 125, 684

\bref
Shetrone, M.~D.,  Siegel, M.~H., Cook, D.~O., \& Bosler, T.\ 2009, AJ, 137, 62 

\bref
Simon, J.~D., \& Geha, M.\ 2007, ApJ, 670, 313 

\bref
Smecker-Hane, T.~A., Stetson, P.~B., Hesser, J.~E.,  \& Lehnert, M.~D.\ 1994, AJ, 108, 507 

\bref
Sohn, S.~T., et al.\ 2007, ApJ, 663, 960 

\bref
Stephens, A., \& Boesgaard, A.~M.\ 2002, AJ, 123, 1647 

\bref
Stoughton, C., et al.\ 2002, AJ, 123, 485 

\bref
Strigari, L.~E., Bullock, J.~S., Kaplinghat, M., Diemand, J., Kuhlen, M., \& Madau, P.\ 2007, ApJ, 669, 676 

\bref
Strigari, L.~E., Bullock, J.~S., Kaplinghat, M., Simon, J.~D., Geha, M., Willman, B., 
\& Walker, M.~G.\ 2008, Nature, 454, 1096 

\bref 
Suntzeff, N.~B., Mateo, M., Terndrup, D.~M., Olszewski, E.~W., Geisler, D., \& Weller, W.\ 1993, ApJ, 418, 208 

\bref
Thielemann, F.-K., et al.\ 2001, PrPNP, 46, 5 

\bref
Tinsley, B.~M.\ 1975, ApJ, 197, 159 

\bref
Tinsley, B.~M.\ 1979, ApJ, 229, 1046 

\bref
Tollerud, E.~J., Bullock, J.~S., Strigari, L.~E., \& Willman, B.\ 2008, ApJ, 688, 277 

\bref
Tolstoy, E., Venn, K.~A., Shetrone, M.~D., Primas, F., Hill, V., Kaufer, A., \& Szeifert, T. 2003, AJ, 125, 707
 
\bref 
 Tolstoy, E., et al.\  2004, ApJL, 617, L119 

\bref
Unavane, M., Wyse, R.~F.~G., \& Gilmore, G.\ 1996, MNRAS, 278, 727 

\bref
van den Bergh, S.\ 2008, MNRAS, 390, L51 

\bref
 Venn, K.~A., Irwin, M.~I., Shetrone, M.~D., Tout, C.~A., Hill, V., \& Tolstoy, E. 2004, AJ, 128, 1177

\bref
Walker, M.~G., Mateo, M., Olszewski, E.~W., Gnedin, O.~Y., Wang, X., Sen, B., 
\& Woodroofe, M.\ 2007, ApJ, 667, L53 

\bref
Walker, M.~G., Mateo, M., Olszewski, E.~W., Sen, B., \& Woodroofe, M.\ 2009, AJ, 137, 3109 

\bref
Walsh, S.~M., Jerjen, H.,  \& Willman, B.\ 2007, ApJL, 662, L83 

\bref
Wang, X., Woodroofe, M., Walker, M.~G., Mateo, M., \& Olszewski, E.\ 2005, ApJ, 626, 145 

\bref
Wilkinson, M.~I., Kleyna, J.~T., Evans, N.~W., Gilmore, G.~F., Irwin, M.~J., 
\& Grebel, E.~K.\ 2004, ApJL, 611, L21 

\bref 
Willman, B., et al.\  2005a, ApJ, 626, L85 

\bref 
Willman, B., et al.\  2005b, AJ, 129, 2692

\bref
Woosley, S.~E., \& Weaver, T.~A.\ 1995, ApJS, 101, 181 

\bref
Zinn, R., \& West, M.~J.\ 1984, ApJS, 55, 45 

\bref 
Zucker, D.~B., et al.\  2004, ApJL, 612, L121 	

\bref 
Zucker, D.~B., et al.\  2006a, ApJL, 650, L41 	

\bref 
Zucker, D.~B., et al.\  2006b, ApJL, 643, L103

\bref 
Zucker, D.~B., et al.\  2007, ApJL, 659, L21 	

}

\vfill

\end{document}